\documentclass[sigconf,10pt]{acmart} 

\newcommand{\paperTitle}{Multi-Tier Buffer Management and Storage System Design
for Non-Volatile Memory} 
\newcommand{\paperKeywords}{Multi-Tier Buffer
Management, Non-Volatile Memory, Prince}
\newcommand{\paperAuthors}{}

\acmYear{} 
\acmConference[]{}{}{}
\acmDOI{}
\acmISBN{}

\usepackage{xspace}
\usepackage{tabularx}
\usepackage{balance}
\usepackage{xcolor}
\usepackage{alltt}

\usepackage{subfig}
\usepackage{pifont}

\hypersetup{
 pdfauthor = {\paperAuthors}, pdftitle = {\paperTitle}, pdfkeywords =
 {\paperKeywords}, pdfborder={ 0 0 0 }
}

\usepackage{float}
\usepackage{amsmath,amsfonts,amssymb,stmaryrd}
\usepackage{booktabs}
\usepackage[end]{algpseudocode}
\usepackage{algorithm}

\usepackage[capitalize,noabbrev,nameinlink]{cleveref}

\newcommand{\squishitemize}{
 \begin{list}{$\bullet$}
  { \setlength{\itemsep}{0pt}
     \setlength{\parsep}{3pt}
     \setlength{\topsep}{3pt}
     \setlength{\partopsep}{0pt}
     \setlength{\leftmargin}{1.95em}
     \setlength{\labelwidth}{1.5em}
     \setlength{\labelsep}{0.5em} } }

\newcounter{Lcount}
\newcommand{\squishlist}{
    \begin{list}{\arabic{Lcount}. }
   { \usecounter{Lcount}
        \setlength{\itemsep}{0pt}
        \setlength{\parsep}{3pt}
        \setlength{\topsep}{3pt}
        \setlength{\partopsep}{0pt}
        \setlength{\leftmargin}{2em}
        \setlength{\labelwidth}{1.5em}
        \setlength{\labelsep}{0.5em} } }

\newcommand{\squishend}{\end{list}}

\definecolor{todo-color}{rgb}{1,0,0}

\definecolor{comment-color}{rgb}{0.25,0.25,0.25}
\newcommand{\codeComment}[1]{\textnormal{\color{comment-color}{\textit{\textbf{\# #1}}}}\unskip}

\DeclareCaptionType{copyrightbox}                                                                                                                                    
\newcommand{\benchTPCC}{TPC-C\xspace}
\newcommand{\benchVoter}{Voter\xspace}
\newcommand{\benchChbenchmark}{CH-benCHmark\xspace}
\newcommand{\benchAuctionmark}{AuctionMark\xspace}

\begin{document}

\newcommand{\mail}[1]{}

\author{Joy Arulraj}
\affiliation{Georgia Institute of Technology}
\email{jarulraj@cc.gatech.edu}
\author{Andrew Pavlo}
\affiliation{Carnegie Mellon University}
\email{pavlo@cs.cmu.edu}
\author{Krishna Teja Malladi}
\affiliation{Samsung}
\email{k.tej@samsung.com}

\title{\paperTitle}

\begin{abstract}

The design of the buffer manager in database management systems (DBMSs)
is influenced by the performance characteristics of volatile
memory (DRAM) and non-volatile storage (e.g., SSD).
The key design assumptions have been that the data must be migrated to DRAM 
for the DBMS to operate on it and that storage is orders of magnitude slower
than DRAM. But the arrival of new non-volatile memory (NVM) technologies that
are nearly as fast as DRAM invalidates these previous assumptions.

This paper presents techniques for managing and designing a multi-tier storage
hierarchy comprising of DRAM, NVM, and SSD.
Our main technical contributions are a multi-tier buffer manager and a storage
system designer that leverage the characteristics of NVM.
We propose a set of optimizations for maximizing the utility of data migration
between different devices in the storage hierarchy. 
We demonstrate that these optimizations have to be tailored based on 
device and workload characteristics. 
Given this, we present a technique for adapting these optimizations to achieve a
near-optimal buffer management policy for an arbitrary workload and storage
hierarchy without requiring any manual tuning.
We finally present a recommendation system for designing a multi-tier storage
hierarchy for a target workload and system cost budget.
Our results show that the NVM-aware buffer manager and storage system designer
improve throughput and reduce system cost across different transaction and
analytical processing workloads.
\end{abstract}

\maketitle

\section{Introduction}
\label{sec:introduction}

The buffer manager in a DBMS provides access to data stored on
non-volatile storage (e.g., SSD) by bringing them into volatile memory (DRAM)
when they are needed.
The canonical approaches for buffer management in DBMSs are predicated on the 
assumptions that (1) the data must be copied to DRAM for the DBMS to operate
on it, and (2) storage is orders of magnitude slower than
DRAM~\cite{dbarchitecture-07,ramakrishnan02,bernstein87}. 
But emerging non-volatile memory (NVM) technologies upend these design
assumptions.

NVM is a broad class of memory technologies, including 
phase-change memory~\cite{intel18,intel17,intel15,raoux08} and
memristors~\cite{hpe17,strukov08}\footnote{Intel is shipping Optane DIMMs that
bring NVM onto the DDR4 memory bus since mid 2018~\cite{intel18}.}. 
NVM devices support low latency reads and writes similar to DRAM, but with
persistent writes and large storage capacity like an SSD. The traditional approaches for buffer management are
incompatible with this new hardware landscape. This stems from two differences
between NVM and canonical storage technologies.
First, to process disk-resident data, the buffer manager must copy it 
to DRAM before the DBMS can perform any operations. In contrast, the CPU can
directly operate on NVM-resident data. Second, NVM shrinks the performance gap
between volatile and non-volatile devices.

In this paper, we present techniques for managing and designing a multi-tier
storage hierarchy comprising of DRAM, NVM, and SSD
\footnote{First-generation NVM devices are expected to be slower (and less
expensive) than DRAM and, at the same time, faster (but more expensive) than
SSD~\cite{intel17}. To maximize performance and minimize cost of the storage
system, NVM will likely co-exist with DRAM and SSD.}.
We propose a set of optimizations for maximizing the utility of data migration
between different devices in the storage hierarchy. These optimizations are
enabled by the introduction of NVM. For example, since the DBMS can directly
operate on NVM-resident data, the buffer manager need not eagerly copy data
from NVM to DRAM. Our results show that such a \textit{lazy data migration} 
technique ensures that only frequently referenced data is promoted to DRAM.

Recent research has focused on optimizing the buffer management policy for a 
particular NVM technology and storage hierarchy.
Renen et al. present a multi-tier buffer manager that eagerly migrates data 
from SSD to DRAM~\cite{renen18}. When a page is evicted from DRAM, the buffer
manager admits it into the NVM buffer based on whether it was recently accessed.
Kwon et al. present a multi-tier file-system that does not cache NVM-resident
data on DRAM and bypasses DRAM while performing synchronous write
operations~\cite{kwon17}.
Although these buffer management policies work well in their target environment, 
they do not generalize to other NVM technologies, storage hierarchies, and
workloads.

We address this problem by introducing a taxonomy for data migration
optimizations that subsumes the specific techniques employed in previous
systems. We illustrate that the buffer management policy must be tailored 
based on device and workload characteristics.
Given this, we make the case for an adaptation mechanism in the buffer manager,
called \textit{adaptive data migration}, that achieves a near-optimal 
buffer management policy for an arbitrary workload and storage hierarchy without
requiring any manual tuning. Prior research on NVM-aware storage management 
has not tackled the problem of designing a multi-tier storage system for a
target workload and system cost
budget~\cite{oukid14,kimura15,eisenman18,renen18,kwon17,arulraj15}.
We present a storage system recommender to address this problem.
In summary, we make the following contributions:

\squishitemize
\item We introduce a taxonomy for NVM-aware data migration optimizations and
present a policy for managing a multi-tier storage hierarchy (\cref{sec:buffer-management}).
\item We introduce an adaptation mechanism in the buffer manager that achieves
a near-optimal policy for an arbitrary workload and storage
hierarchy without requiring any manual tuning (\cref{sec:adaptive-migration}).
\item We introduce a recommendation system for designing a multi-tier storage
hierarchy for a target workload and system cost
budget (\cref{sec:hierarchy-selection}).
\item We demonstrate that the NVM-aware buffer manager and storage system designer 
improve throughput and reduce cost across different
transaction and analytical processing workloads (\cref{sec:evaluation}).
\squishend

\section{Background}
\label{sec:background}

We now provide an overview of buffer management in DBMSs. 
We then make the case for the introduction of NVM in the storage hierarchy.

\subsection{Buffer Management}
\label{sec:background:buffer-management}

The buffer manager partitions the available memory into a set of fixed-size
slots, which is collectively termed as a \textit{buffer}.
The higher-level components of the DBMS, such as the query execution engine, 
need not concern themselves with whether a page is in the buffer  
or not. They only need to request the buffer manager to retrieve a page.
If a page requested by another component is not present in the buffer, 
the buffer manager transparently retrieves the page from non-volatile storage.

The buffer manager maintains transient meta-data about each page in the
in-memory buffer.
This meta-data includes the number of active references made to the page and
whether the page has been modified since it was brought into the buffer 
from storage. If a page requested by another component is already present in the
buffer, then it increments the number of active references to the page and 
returns the address of the slot containing the page. Otherwise, the buffer
manager chooses a slot for replacement based on the replacement 
policy (e.g., least recently used)~\cite{oneil93}. If the page selected for
replacement contains any modifications, the buffer manager propagates those
changes to the corresponding  page on non-volatile storage. It then copies 
the requested page from storage into the replacement slot and returns the 
slot's address.

The buffer manager does not have complete autonomy over when and what pages are
flushed to non-volatile storage~\cite{agrawal89,franklin97}.
It coordinates with the DBMS's \textit{log manager} to ensure that the 
changes made by a transaction are durable when it is committed, 
and that the changes made by transactions that were not committed at
the time of a system failure are reversed during recovery. These constraints are 
referred to as the \textit{durability} and \textit{failure atomicity}
properties. 

If a transaction modifies a block and then commits, and the buffer manager has
not yet written the updated block to durable storage, then a failure
will leave the block in its old invalid state, thereby violating the durability
property. On the other hand, if the buffer manager decides to write a 
modified block belonging to an active transaction, it violates the
atomicity property. To prevent such scenarios, the buffer manager refrains
from making autonomous replacement decisions.

Since the contents of the DRAM buffer are lost after a system failure, 
the log manager records information needed to recover from a failure on
durable storage. Before updating a page, the DBMS writes its old contents 
to the log (i.e., the before image of the page). 
Similarly, when a page is about to be evicted from the buffer pool, its current
contents are recorded in the log (i.e., the after image of the page). 
During recovery, the DBMS uses the information in the log to restore the
database to a transactionally consistent state.
To bound the amount of time taken to replay the log during recovery, 
the DBMS periodically takes checkpoints at runtime~\cite{mohan92}. 


\begin{table}[t!]
    \centering
    {\small {
\setlength{\tabcolsep}{3pt}
\begin{tabular}{ll|ll|ll}	
	\toprule
                            & \textbf{DRAM} & \textbf{PCM} & \textbf{RRAM}
                            & \textbf{SSD} &                       
                            \textbf{HDD}  \\
    \midrule
    Read latency            &  50 ns & 50 ns  & 100 ns 
                            & 25 $\mu$s & 10 ms \\
    Write latency           &  50 ns & 200 ns & 100 ns 
                            & 300 $\mu$s & 10 ms \\
    Bandwidth               &  60 GB/s & 10 GB/s & 10 GB/s 
                            & 1 GB/s & 0.1 GB/s \\ 
	\$/GB                   &  10  &  1 &  1 
	                        &  0.2 & 0.02\\
    Persistent              &  No   & Yes     & Yes     
                            & Yes    & Yes 
    \\
    Endurance               & $>\!\!10^{16}$ & $10^{10}$ 
                            & $10^{15}$ & $10^{5}$ & $>\!\!10^{16}$ \\
    \bottomrule
\end{tabular}
}}
    \caption{
        Comparison of candidate NVM technologies~\cite{chen14,dulloor14,perez10,moraru13}:
        phase-change memory (PCM)~\cite{intel17,intel15,raoux08} and 
        memristors (RRAM)~\cite{hpe17,strukov08}. The
        price of NVM is derived from the current price of Intel's 3D XPoint-based 
        Optane SSD 900P~\cite{optane18}.
    }
    \label{tab:comparison}
\end{table}

\subsection{Non-Volatile Memory DBMSs}
\label{sec:background:nvm}

\begin{figure}
    \centering    
    \subfloat[DRAM-SSD Hierarchy]{
        \includegraphics[width=0.22\textwidth]{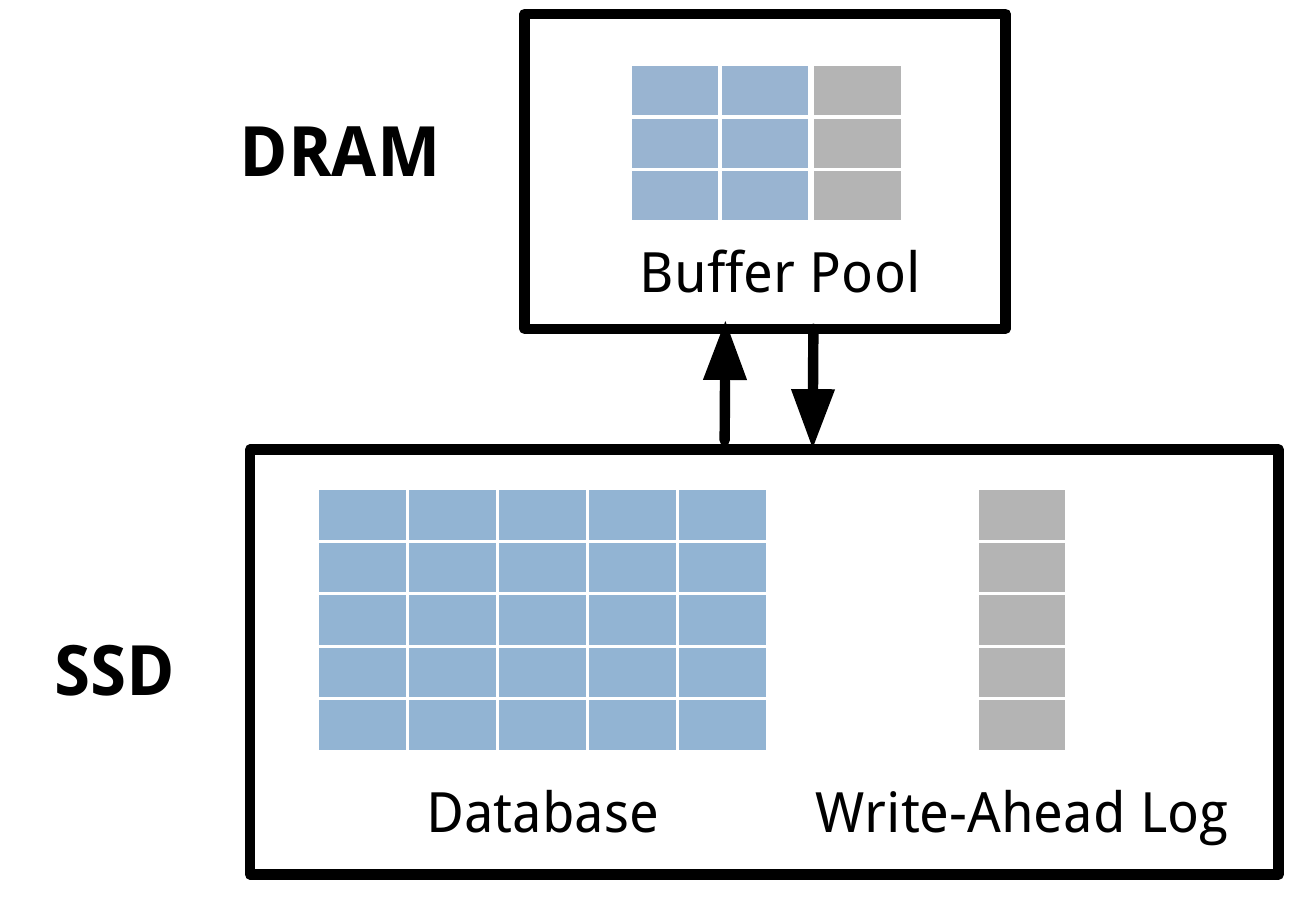}
        \label{fig:arch-1}
    }    
    \hfill
    \subfloat[NVM-SSD Hierarchy]{
        \includegraphics[width=0.22\textwidth]{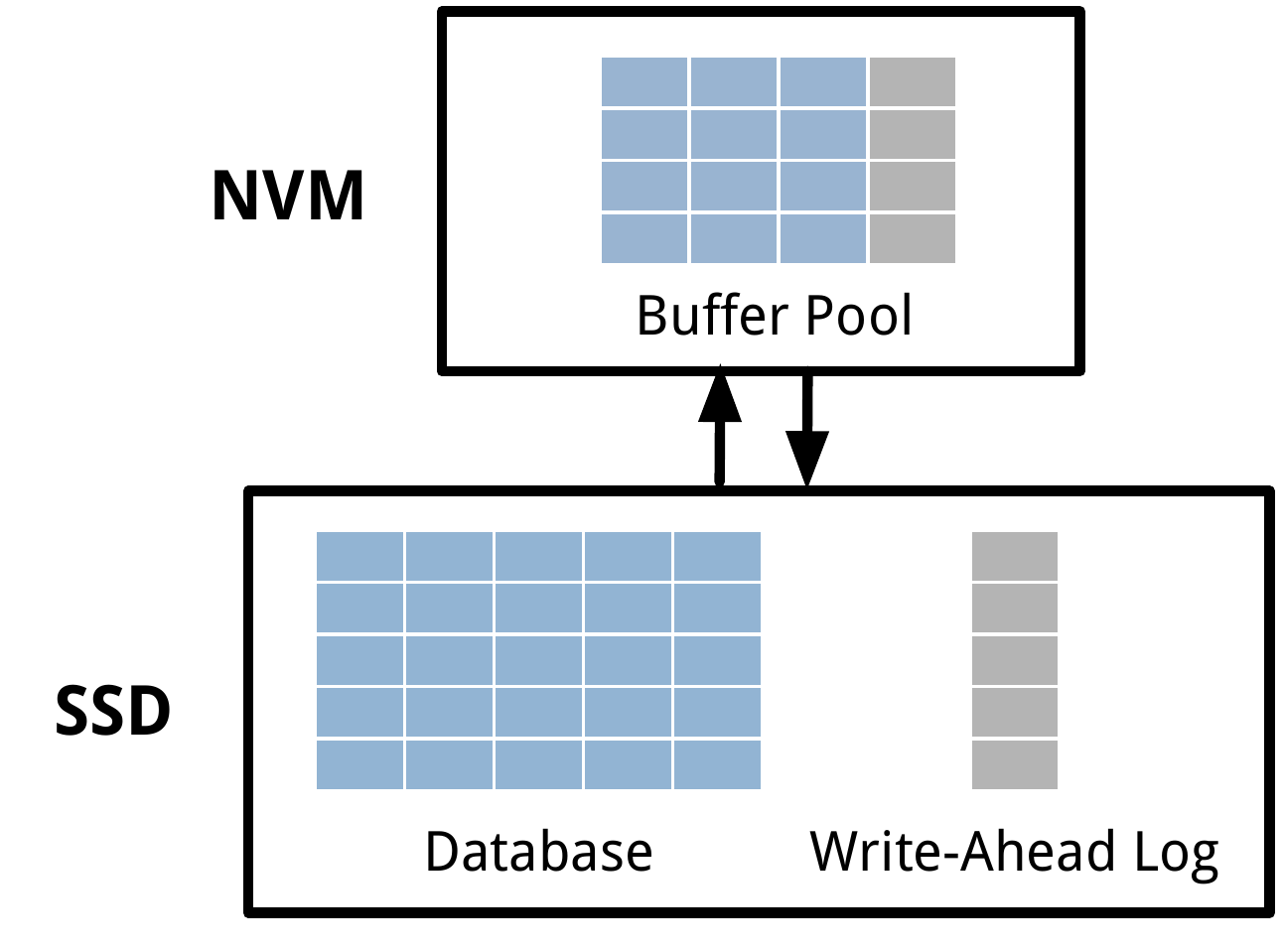}
        \label{fig:arch-2}
    }    
    \hfill
    \subfloat[DRAM-NVM-SSD Hierarchy]{
        \includegraphics[width=0.22\textwidth]{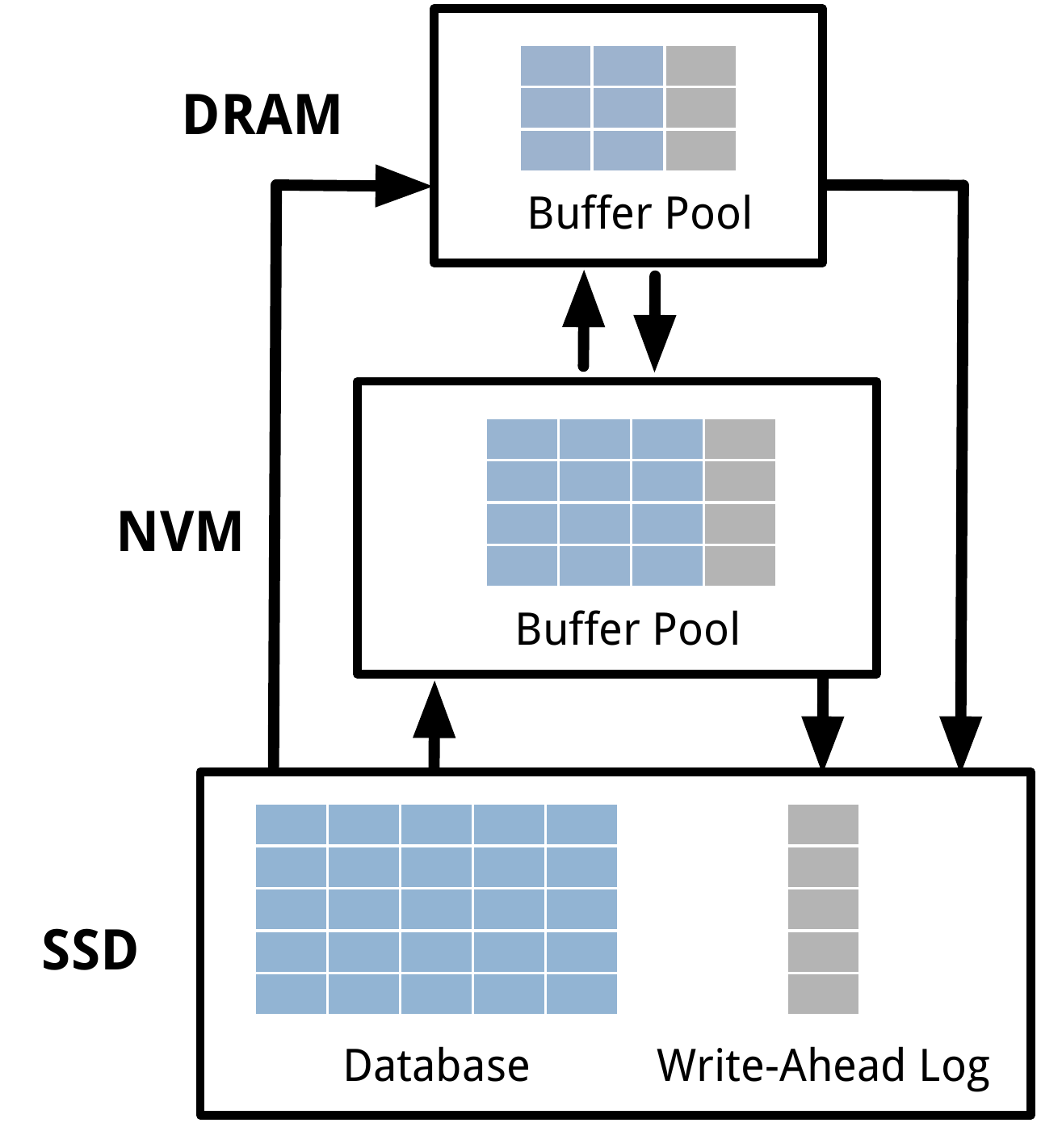}
        \label{fig:arch-3}
    }    
    \caption{
        \textbf{Storage Hierarchies} -- Candidate storage hierarchies: 
		(a) DRAM-SSD, (b) NVM-SSD, and 
	    (c) DRAM-NVM-SSD.
    }
    \label{fig:arch}
\end{figure}

A DBMS's performance is constrained by the speed with which it can retrieve
data from and persist data (e.g., pages containing log records) on
disk~\cite{harizopoulos08}. As illustrated in~\cref{fig:arch-1}, 
the buffer manager copies pages from SSD to DRAM for the DBMS to operate on
them. Since DRAM accesses are 100$\times$ faster than SSD operations,
DBMSs manage a large buffer pool on DRAM. 
It is difficult, however, to  deploy high-capacity DRAM systems due to three
factors. First, it drives up the total cost of the system since it is
50$\times$ more expensive than secondary storage technologies. 
Second, increasing DRAM capacity raises the total system
power consumption. Lefurgy et al. report that as much as 40\% of the total
system energy is consumed by DRAM in commercial servers~\cite{lefurgy03}.
Lastly, DRAM scaling faces significant challenges due to limitations in
scaling techniques used in earlier generations for transistors and 
capacitors~\cite{mandelman02}.

Emerging NVM technologies bridge the performance and cost differentials
between DRAM and SSD. \cref{tab:comparison} compares the characteristics
of candidate NVM technologies. NVM latency is within an order of magnitude
higher than that of DRAM. Unlike SSDs/HDDs that use the SAS or SATA interfaces,
NVM can be plugged into DIMM slots to deliver higher bandwidths and lower
latencies to CPUs.

NVM supports higher data density than DRAM due to its intrinsic device
characteristics\footnote{For example, phase-change memory (PCM) is a
NVM technology that exploits the ability of chalcogenide glass to oscillate 
between amorphous and crystalline states when heated using electrical
pulses~\cite{raoux08}. A PCM cell can exist in different degrees of partial
crystallization, thereby enabling more than one bit to be stored in each cell.}. 
It is, therefore, less expensive than DRAM. Similar to SSDs, the number of write
cycles per bit is limited in NVM technologies.
Reducing the number of writes using an NVM-aware buffer management policy
extends the lifetime of devices with limited write-endurance.

Since NVM devices are slower than DRAM, replacing the latter with a
similarly priced NVM device with higher capacity can reduce the performance of
the DBMS. This architecture, as illustrated in ~\cref{fig:arch-2}, delivers
performance comparable to that of a DRAM-SSD hierarchy only if NVM latency is 
less than 2$\times$ that of DRAM (\cref{sec:exps-size}).
 
A multi-tier storage hierarchy with DRAM, NVM, and SSD, as shown in
~\cref{fig:arch-3}, can simultaneously maximize performance and minimize cost of
the storage system. The reasons for this are twofold. First, the NVM buffer
caches a significant fraction of the working set, thereby reducing SSD accesses.
Second, the DRAM buffer serves as a cache on top of NVM and only stores the
hottest pages in the database.

In a DRAM-SSD hierarchy, the buffer manager decides \textit{what} pages
to move between disk and memory and \textit{when} to move them.
With a DRAM-NVM-SSD system, however, in addition to deciding 
what/when data should be migrated, it must also decide \textit{where} to move
them (i.e., what storage tier). In the next section, we discuss how this
decision is influenced by the characteristics of NVM.

\section{NVM-Aware Buffer Management}
\label{sec:buffer-management}

NVM introduces new data flow paths in the storage hierarchy. 
By leveraging these additional options, the buffer manager reduces 
data movement between different tiers and minimizes the number of 
writes to NVM.
The former results in improving the DBMS's performance, while the latter 
extends the lifetime of NVM devices with limited write-endurance~\cite{raoux08}.

\begin{figure}
    \centering    
    \includegraphics[width=0.45\textwidth]{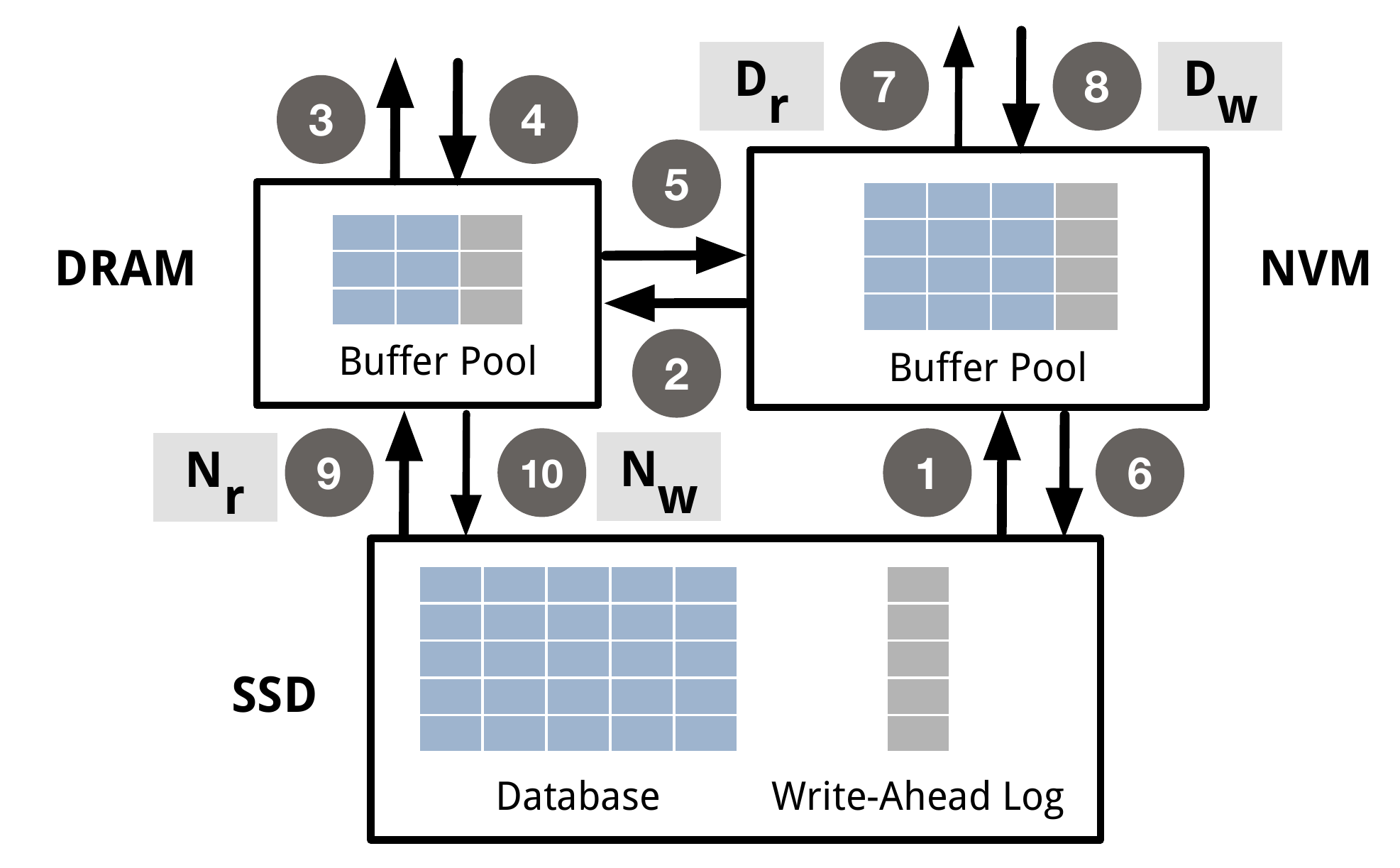}
    \caption{
        \textbf{Data Flow Paths} --
       	The different data flow paths in a multi-tier storage hierarchy
       	consisting of DRAM, NVM, and SSD.
    }
    \label{fig:data-flow}
\end{figure}

\cref{fig:data-flow} presents the data flow paths in the multi-tier storage
hierarchy. The default read path comprises of three steps:
moving data from SSD to NVM (\ding{202}), then to DRAM (\ding{203}), and lastly
to the processor cache (\ding{204}). Similarly, the default write path consists
of three steps: moving data from processor cache to DRAM (\ding{205}), 
then to NVM (\ding{206}), and finally to SSD (\ding{207}). 
We now describe how the buffer manager leverages the additional data flow paths
in~\cref{fig:data-flow} (\ding{208},\ding{209},\ding{210},\ding{211}) to
minimize the performance impact of NVM and to extend the lifetime of the NVM
device.

\subsection{Bypass DRAM during Reads}
\label{sec:data-flow-1}

Unlike SSDs, the processor can directly access data on NVM via read operations
(\ding{208}). To access a block on SSD, in a disk-centric DBMS, the DBMS copies 
it over to DRAM (\ding{210}), before it can operate on the copied data. 
With NVM, the buffer manager can leverage this new data flow path to lazily
migrate data from NVM to DRAM while serving read operations.

Let $\mathcal{D}_{r}$ represent the \textit{probability} that the buffer
manager copies data to DRAM during read operations. With existing
storage technologies, $\mathcal{D}_{r}$ = 1. We refer to this data migration
policy as \textit{eager migration}. With NVM, the buffer manager can employ 
a wider range of \textit{lazy migration} policies with smaller values for 
$\mathcal{D}_{r}$.
Such policies reduce upward data migration between NVM and DRAM
during read operations. They are beneficial when the capacity of DRAM is 
smaller than that of NVM. A lazy migration strategy ensures that colder data
(i.e., data that has not been frequently accessed) on NVM do not evict hotter
data in DRAM. 

The optimal value of $\mathcal{D}_{r}$ depends on the application's workload.
An eager migration policy ($\mathcal{D}_{r}$ $\leq$ 0.5) works well if the
working set fits within the DRAM buffer.
A lazier policy would delay the promotion of data to DRAM, thereby
increasing the impact of NVM latency on performance.
However, a migration policy with higher $\mathcal{D}_{r}$ works well 
if the working set does not fit in DRAM and fits in the NVM buffer. 
This strategy ensures that only the hot data is stored in DRAM.

In addition to the size of the working set, $\mathcal{D}_{r}$ depends on the
ratio between the capacities of the DRAM and NVM buffers. 
In a storage hierarchy where the ratio approaches one, a more eager policy
enables the buffer manager to leverage the space available in DRAM. 
Smaller values for $\mathcal{D}_{r}$ work well when the ratio approaches zero
since they ensure that the DRAM buffer only contains frequently referenced
data.

With the eager migration policy, the buffer manager always brings the block to
DRAM while serving the read operation. Consequently, if the application then 
updates the same block, the writes are performed on DRAM. 
In contrast, a lazy migration policy increases the number of writes on NVM. 
This is because it is more likely that the block being updated is residing on
NVM when the buffer manager adopts such a policy. This is not a problem for
DBMS applications with skewed access patterns~\cite{beaver10,sikka12}.
Such applications tend to modify hot data that is cached in DRAM even when the
buffer manager employs a lazy migration policy.

\subsection{Bypass DRAM during Writes}
\label{sec:data-flow-2}

Ensuring the persistence of pages containing log and checkpoint records is
critical for the recoverability of the DBMS as we discussed
in~\cref{sec:background:buffer-management}.
The DBMS's performance is constrained by the I/O overhead associated with
persisting these pages on non-volatile storage.  
As transactions tend to generate  multiple log records that are each small in
size, most DBMSs use the \textit{group commit} optimization to reduce this 
I/O overhead~\cite{dewitt84}. 
The DBMS first batches the log records for a group of transactions in the DRAM
buffer (\ding{205}) and then flushes them together with a single write to SSD
(\ding{207}). This improves the operational throughput and amortizes the I/O 
overhead across multiple transactions.

Unlike SSDs, the CPU can directly persist data on NVM via write operations
(\ding{209}).
The buffer manager leverages this path to provide \textit{synchronous
persistence} with lower overhead~\cite{arulraj17b,kwon17}.
The write operation bypasses DRAM since the data must be eventually
persisted, and this data migration optimization shrinks the overall latency of
the operation, especially on NVM devices whose write latencies are comparable to
that of DRAM. In addition to eliminating the redundant write to DRAM, 
it also avoids potential eviction of other hot pages from the DRAM buffer.

Let $\mathcal{D}_{w}$ represent the probability with which the buffer
manager copies data into DRAM during write operations. 
With canonical storage technologies, similar to $\mathcal{D}_{r}$,
$\mathcal{D}_{w}$ = 1. With NVM, the buffer manager can employ lazy migration 
policies with smaller $\mathcal{D}_{w}$.
Such policies reduce the frequency of downward data migration to DRAM 
during write operations thereby ensuring that pages containing log and
checkpoint records do not evict hotter data in DRAM.

\subsection{Bypass NVM During Reads}
\label{sec:data-flow-3}

The data migration optimizations presented in
~\cref{sec:data-flow-1,sec:data-flow-2} improve the performance of the DBMS 
at the expense of increasing the number of writes to NVM.
We next present optimizations for reducing the number of writes to
NVM using alternate data flow paths.

The default read path consists of moving the data from SSD to NVM (\ding{202})
and eventually migrating it to DRAM (\ding{203}). This optimization makes 
use of the direct data flow path from SSD to DRAM (\ding{210}).
When the buffer manager observes that a requested page is not present in
both the DRAM and NVM buffers, it copies the data on SSD directly to DRAM, 
thus bypassing NVM during read operations. If the data read into the DRAM buffer
is not subsequently modified, and is selected for replacement, 
then the buffer manager discards it. If the page is modified and later selected
for eviction from DRAM, the buffer manager considers admitting it to NVM
(\ding{206}).

Let $\mathcal{N}_{r}$ represent the probability with which the buffer
manager copies data from SSD to NVM during read operations.
With the default read path, $\mathcal{N}_{r}$ = 1. 
When a page is fetched from SSD and later evicted from DRAM, 
an eager policy necessitates two writes to NVM: once at fetch time and again
when the page is evicted from DRAM. With a lazy policy (i.e., smaller
$\mathcal{N}_{r}$), the buffer manager installs a copy of a modified page on NVM
only after it has been evicted from DRAM. This eliminates the first write to NVM
when the page is fetched from SSD.

\subsection{Bypass NVM During Writes}
\label{sec:data-flow-4}

Another data migration optimization for reducing the number of writes to NVM
consists of bypassing NVM while serving write operations.
The default write path consists of moving the data from DRAM to NVM (\ding{206})
and then eventually migrating it to SSD (\ding{207}).  
Instead of using the default path, this optimization makes use of the direct
data flow path from DRAM to SSD (\ding{211}).

By bypassing NVM during writes, the buffer manager ensures that only pages
frequently swapped out of DRAM are stored on NVM~\cite{renen18}.
This optimization reduces the number of writes to NVM since only warmer pages
identified by the buffer manager are stored in the NVM buffer. 
If the buffer manager employs an eager policy while reading data into
DRAM (i.e., high $\mathcal{D}_{r}$), this optimization prevents colder 
DRAM-resident pages from polluting the NVM buffer.

Let $\mathcal{N}_{w}$ represent the probability with which the buffer manager
copies data from DRAM to NVM during write operations.
With the default write path, $\mathcal{N}_{w}$ = 1.  
Lower values of $\mathcal{N}_{w}$ reduce downward data migration into NVM. 
Such a lazy policy is beneficial when the capacity of DRAM is comparable to that
of NVM since it ensures that colder data on DRAM does not evict warmer
data in the NVM buffer.

\section{Adaptive Data Migration}
\label{sec:adaptive-migration}

The \textit{data migration policy} consists of the probabilities with which the
buffer manager bypasses DRAM and NVM while serving read and write operations
(\cref{sec:data-flow-1,sec:data-flow-2,sec:data-flow-3,sec:data-flow-4}).
All of the above data migration optimizations are moot unless the buffer manager
can effectively adapt the overall policy based on the characteristics
of the workload and the storage hierarchy. 
We now present an adaptation mechanism in the buffer manager that achieves a
near-optimal policy for an arbitrary workload and storage hierarchy without
requiring any manual tuning.
 
The crux of our approach is to track the \textit{target metrics} on recent query
workload at runtime, and then periodically adapt the  policy in the background.
Over time, this process automatically optimizes the policy for the application's
workload and the storage hierarchy, and amortizes the adaptation cost 
across multiple queries. We now describe the information that the buffer manager
collects to guide this process.

The buffer manager keeps track of two target metrics while executing the
workload. These include the operational throughput of the buffer manager and the
number of write operations performed on NVM. The goal is to determine the
optimal configuration of the data migration policies that maximizes the
throughput and minimizes writes to NVM. The \textit{cost function} associated
with a candidate data migration policy configuration consists of two weighted
components associated with these target metrics.
$$Cost(T, W) = T + (\lambda/W)$$


To adapt the buffer manager's data migration policy, we employ an iterative
search method called \textit{simulated annealing} (SA)~\cite{kirkpatrick83}. 
This technique searches for 
a policy configuration that maximizes the cost function presented.
An attractive feature of SA is that it avoids getting caught at local optima,
which are configurations that are better than any other nearby configurations,
but are not the globally optimal configuration~\cite{ioannidis87}. 
It is a probabilistic hill climbing algorithm that migrates through a
set of local optima in search of the global extremum. 

SA consists of two stochastic processes for generating candidate policy
configurations and for accepting a new configuration.
~\cref{alg:sa} presents the algorithm for tuning the data migration policy using
SA. At each time step, SA randomly selects a new configuration
($C'$) close to the current one ($C$). It then evaluates the cost of that
configuration ($E'$). Lastly, it decides to accept the configuration $C'$ 
or stay with $C$ based on whether the cost of $C'$ is lower or higher than that
of the current configuration. If $C'$ is better than $C$, then it immediately
transitions to $C'$. Otherwise, it randomly accepts the new configuration with
higher cost ($C'$) based on the Boltzmann acceptance probability factor.

\begin{algorithm}[t!]
    \small
    \caption{Data Migration Policy Tuning Algorithm}
    \label{alg:sa}
    \begin{algorithmic}
    \Require
    temperature reduction parameter $\alpha$,\\
    threshold for number of accepted transitions $\gamma$,\\
    initial data policy configuration $C_{0}$,\\   
    initial temperature $T_{0}$,\\   
    final temperature $T_{min}$   

    \Function{UPDATE-CONFIGURATION}{$\alpha,\gamma,C_{0},T_{0},T_{min}$} 
    \State \codeComment{Initialization}
	\State current configuration $C$ = $C_{0}$
	\State energy $E$ = cost($C$)
	\State temperature $T$ = $T_{0}$
    \State \codeComment{Iterative Search}     
    \While{$T$ > $T_{min}$}
    	\While {number of accepted transitions < $\gamma$}
    		\State new configuration $C'$ = neighbor($C$)
    		\State energy $E'$ = cost($C'$)    	
    		\State energy delta $\Delta E$ = $E'$ -$E$
    		\State Boltzmann acceptance probability $P$ = $e^{\frac{- \Delta E}{T}}$
			\If{$\Delta E$ < 0 or with acceptance probability $P$}				
				\State \codeComment{Accept new policy configuration}
	            \State $C$ = $C'$ 
	        \EndIf
	    \EndWhile
	    \State \codeComment{Reduce temperature}
	    \State $T$ = $T$ * $\alpha$
    \EndWhile
\EndFunction  
\end{algorithmic}
\end{algorithm}  

SA is theoretically guaranteed to reach the global optima with high probability.
The control parameter $T$ determines the magnitude of the perturbations of the
energy function $E$. SA gradually decreases $T$ over time. 
During the initial steps of SA, at high temperatures, the probability of uphill 
moves in the energy function ($\Delta$E > 0) is large. Despite temporarily
increasing the energy, such non-beneficial downhill steps ($\Delta$E < 0) 
allows for a more extensive search for the global optimal configuration.
Over time, SA reduces the temperature. This gradual cooling mechanism
corresponds to slowly decreasing the probability of accepting worse 
configurations as it explores the configuration state space.

\section{Storage Hierarchy Selection}
\label{sec:hierarchy-selection}

We have so far focused on identifying an optimal data migration policy
configuration for a particular workload given a storage
hierarchy. The tuning algorithm presented in~\cref{sec:adaptive-migration}
assumes that we have already provisioned a multi-tier storage hierarchy that is
a good fit for the workload. It is unclear, however, how to select such a 
hierarchy for a particular workload given a system cost constraint. 

In this section, we formulate an analytical model of a hierarchical
storage system to improve the intuition behind provisioning a multi-tier storage
hierarchy. We then identify the limitation of the model and present a
recommender system that addresses them.

\subsection{Hierarchical Storage System Model}
\label{sec:model}

We can model the multi-tier storage system as a linear hierarchy with $n$
levels, $L_{1}$, $L_{2}$,$\ldots$, $L_{n}$. The performance of a particular
level $L_{i}$ in the hierarchy is determined by two factors: 
the average access time $t_{i}$ and the device capacity
$C_{i}$~\cite{jacob96}. We assume that a copy of all blocks in level $i$
exists in every level greater than $i$ (i.e., in all lower levels in the hierarchy).
The maximum information that can be stored in the system is equal to the
capacity of the lowest level $C_{n}$, since copies of all blocks stored in the
higher levels of the system must be present in $L_{n}$.

We can characterize the performance impact of the device capacity at a
particular level by the probability of finding the requested data block in that
level. This is termed as the \textit{hit ratio} $H$. $H$ is a
monotonically increasing function with respect to device capacity $C$.
Let the cost per storage unit (e.g., per GB) of the device
technology used at a particular level be given by the \textit{cost
function} $P(t_{i})$. It decreases monotonically with respect to the access
time $t_{i}$ of the device technology.

Since a copy of all data blocks at level $i$ exists in every level greater than
$i$, the probability of a hit in level $L_{i}$ and misses in the higher levels, 
is given by:
$$h_{i} = H(C_{i}) - H(C_{i-1})$$

Here, $h_{i}$ represents the relative number of successful data accesses at
level $i$ in the storage hierarchy. The \textit{effective average access time}
per block request, is then given by:
$$T = \sum_{i=1}^{n} h_{i} (\sum_{j=1}^{i} t_{j})$$

To maximize the operational throughput of the DBMS, we need to minimize 
$T$ subject to storage system cost constraints. Given a storage system 
cost budget $B$, the goal is to select the device technology $t_{i}$ and
determine the device capacity $C_{i}$ for each level in the storage hierarchy.
We formulate this problem as follows:\\

Minimize: 
$$T = \sum_{i=1}^{n} (1-H(C_{i-1}))t_{i}.$$

Subject to the storage system cost budget:
$$\sum_{i=1}^{n} P(t_{i}) C_{i} \leq B$$

\subsection{Storage Hierarchy Recommender System}
\label{sec:recommender}

$H$ is a function of the workload locality and does not have a closed-form
expression. We circumvent this limitation by developing a recommender system 
that measures the actual throughput on a target workload across 
candidate storage hierarchies to identify the optimal system.
The goal of the recommender system is to identify a multi-tier storage
hierarchy consisting of DRAM, NVM, and/or SSD that maximizes a user-defined
objective function given a system cost budget. 

The recommender system searches across candidate storage hierarchies that 
meet the user-specified budget. Let $\{D_{0},\\ D_{1}, D_{2},\ldots, D_{p}\}$
represent the set of candidate DRAM devices,
$\{N_{0}, N_{1}, N_{2},\ldots, N_{q}\}$ the set of candidate NVM devices, 
and $\{S_{0}, S_{1}, S_{2},\ldots, S_{r}\}$ the set of candidate SSD devices.
These devices have varying capacities and costs. 
We are provided with a cost function $P$ that returns the cost of a particular
device. For instance, $P(D_{i})$ returns the cost of the DRAM device with
capacity $D_{i}$.

We can prune the set of candidate storage hierarchies by only considering
devices whose capacities are powers of two. With this restriction, 
the size of set of candidate storage hierarchies is small ($p$, $q$,
and $r$ < 10). The recommender system does a \textit{pure grid search} over
the entire set~\cite{bergstra12}. During a particular trial on a grid, we only
consider device triples $\{D_{i}, N_{j}, S_{k}\}$ that meet the user-specified
budget $B$, as given by:
$$P(D_{i}) + P(N_{j}) + P(S_{k}) \leq B$$

The system then measures the operational throughput on the storage hierarchy
corresponding to the device triple $\{D_{i}, N_{j}, S_{k}\}$.
We configure $D_{0}=0$ to model storage hierarchies containing only NVM
and SSD devices (i.e., those that do not have DRAM). Similarly, we set $N_{0}=0$
and $S_{0}=0$ to model storage hierarchies without NVM and SSD, respectively. 
We note that the entire database must fit in the lowest level of storage
hierarchy. Since the cost of NVM is 10$\times$ higher than that of SSD,
the latter device will likely continue to occupy the lowest level.

\section{Experimental Evaluation}
\label{sec:evaluation}

In this section, we present an analysis of the proposed NVM-aware buffer
management policies and the storage hierarchy recommendation system. 
Our goal is to demonstrate that:

\squishitemize
\item NVM improves throughput by reducing accesses to canonical storage devices 
due to its higher capacity-cost ratio compared to DRAM 
(\cref{sec:runtime-performance}).
\item The selection of a data migration policy depends on the runtime
performance requirements, write endurance characteristics of NVM, and the
relative size of the DRAM buffer compared to NVM (\cref{sec:exps-migration}).
\item Tuning the buffer management policy for the workload and the storage
hierarchy improves throughput and extends the lifetime of the NVM device
(\cref{sec:exps-tuning}).
\item The selection of a multi-tier storage hierarchy for a given workload
depends on the working set size, the frequency of persistent writes, the
system cost budget, and the performance and cost characteristics of
NVM (\cref{sec:exps-size}).
\item A combination of data migration optimizations presented
in~\cref{sec:buffer-management} outperforms the state-of-the-art buffer
management policy (\cref{sec:exps-comparison}).
\squishend

\subsection{Trace-Driven Buffer Manager}
\label{sec:trace-driven-buffer-manager}

We developed a trace-driven buffer manager to evaluate different storage
hierarchy designs and data migration policies. We gather traces from a
real DBMS by running OLTP, OLAP, and HTAP workloads. 
The trace contains information about individual buffer pool operations.

At the beginning of the trace period, we take a snapshot of the DBMS's meta-data
regarding the blocks stored in memory and on storage. This snapshot does
not contain any user data. The buffer manager only simulates the movement of
user data blocks and not their actual contents. This allows us to effectively
run simulations of buffer management operations on large databases and devices.

The buffer manager runs on top of a multi-tier storage hierarchy
consisting of DRAM, NVM, and/or SSD. For instance, in case of a three-tier 
DRAM-NVM-SSD hierarchy, it maintains two buffer pools on DRAM and NVM.
While processing the trace requests, the buffer manager issues read and write
operations to the appropriate devices depending on the data migration policy.
The simulator models the physical contiguity of the user-data blocks while
distributing the I/O operations across the device.

We conduct our experiments on a NVM hardware emulator. Existing NVM devices
cannot store large databases due to their limited capacities and 
prohibitive costs. We instead use the persistent memory evaluation platform 
(PMEP) developed by Intel Labs~\cite{dulloor14,zhang15}. PMEP models the
latency and bandwidth characteristics of upcoming NVM technologies. 
It allows us to tune the memory read and write latencies and bandwidths. 
This enables us to evaluate multiple NVM device profiles that are not specific
to a particular technology. A detailed description of PMEP is provided
in~\cref{sec:emulator}.

\subsection{Experimental Setup}

We perform our experiments by running the trace-driven buffer manager on the NVM
hardware emulator. By default, we set the capacity of the DRAM and NVM buffers 
to be 2~GB and 128~GB, respectively. Unless otherwise stated, we configured the
NVM latency to be 2$\times$ that of DRAM and validated these settings using 
Intel's memory latency checker. The emulator's storage hierarchy also includes
two additional devices:

\squishitemize
\item \textbf{HDD:} Seagate Barracuda (3 TB, 7200 RPM, SATA 3.0)
\item \textbf{SSD:} Intel DC S3700 (400 GB, SATA 2.6)
\squishend

\textbf{Workloads}
We next describe the workloads from the OLTP-Bench testbed that we use in our
evaluation~\cite{oltpbench,difallah13}. These workloads differ in their workload
skews and frequencies of persistent writes.
 \\ \vspace{-0.05in}
 
\textbf{\benchTPCC:}
This benchmark is the industry standard for evaluating the performance of OLTP
systems~\cite{tpc-c}. It simulates an order-entry application of a wholesale
supplier and consists of five transaction types with nine tables.
\\ \vspace{-0.05in}

\textbf{\benchVoter:}
This is an OLTP benchmark that simulates a phone-based election application. It
is derived from the software system used to record votes for a television talent show.
The workload consists of short-lived transactions that each update a small
number of tuples. 
\\ \vspace{-0.05in}

\textbf{\benchChbenchmark:}
This is a complex HTAP workload that is derived from a transactional workload 
based on the order entry processing of TPC-C and a corresponding
TPC-H-equivalent OLAP query suite. It is useful to evaluate DBMSs designed to
serve both OLTP and OLAP  workloads. \benchChbenchmark extends the \benchTPCC
benchmark with 22 additional analytical queries.
\\ \vspace{-0.05in}

\textbf{\benchAuctionmark:}
This is an OLTP benchmark that models the workload characteristics of an on-line
auction site~\cite{angkanawaraphan17}. 
The user-to-item ratio follows a highly skewed Zipfian distribution. 
The total number of transactions that target each item is temporally skewed, as
items receive more activity as the auction approaches its closing.
\\ \vspace{-0.05in}

\textbf{Trace Collection}
\label{sec:trace-collection}
We collect traces by running the benchmarks on an instrumented fork of Postgres
DBMS (v9.4)~\cite{postgres}.
All the transactions execute with the same serializable isolation level and
durability guarantees. To collect the traces, we first ran each benchmark for a
warm-up period.
At the end of the warm-up period, we take a snapshot of the DBMS's meta-data
regarding the location of blocks in volatile memory and on durable storage. 
We then start recording the buffer pool references in the trace. During
simulation, the buffer manager first loads the snapshot before executing the 
operations recorded in the trace.

The amount of data referenced at least once in a trace is termed as its
\textit{footprint}. An important issue in using trace-driven
simulations to study storage hierarchy design is that the traces must have
a sufficiently large footprint for the storage configurations of
interest~\cite{hsu01}. ~\cref{tab:trace} presents the footprints of the traces
associated with different benchmarks. For all experiments, we used half of 
the trace to warm-up the simulator. We collect system statistics only after the
buffer pools have been warmed up.

\begin{table}[t!]
    \centering
    {\small {

\newcolumntype{b}{X}
\newcolumntype{Y}{>{\centering\arraybackslash}X}
\begin{tabularx}{\columnwidth}{YY}
\toprule
\textbf{Benchmark} & \textbf{Footprint}\\

\midrule

\benchTPCC & 1.32~TB  \\ 
\benchChbenchmark & 1.13~TB  \\ 
\benchVoter & 1.05~TB  \\ 
\benchAuctionmark  & 815~GB \\

\midrule

\end{tabularx}
}}
    \caption{
        \textbf{Trace Footprints:} Footprints of the traces
        associated with different benchmarks.
    }
    \label{tab:trace}
\end{table}

\subsection{Workload Skew Characterization}
\label{sec:workload-characterization}

\begin{figure}[t!]
    \centering
    \subfloat[TPCC]{
        \includegraphics[width=0.22\textwidth]
                        {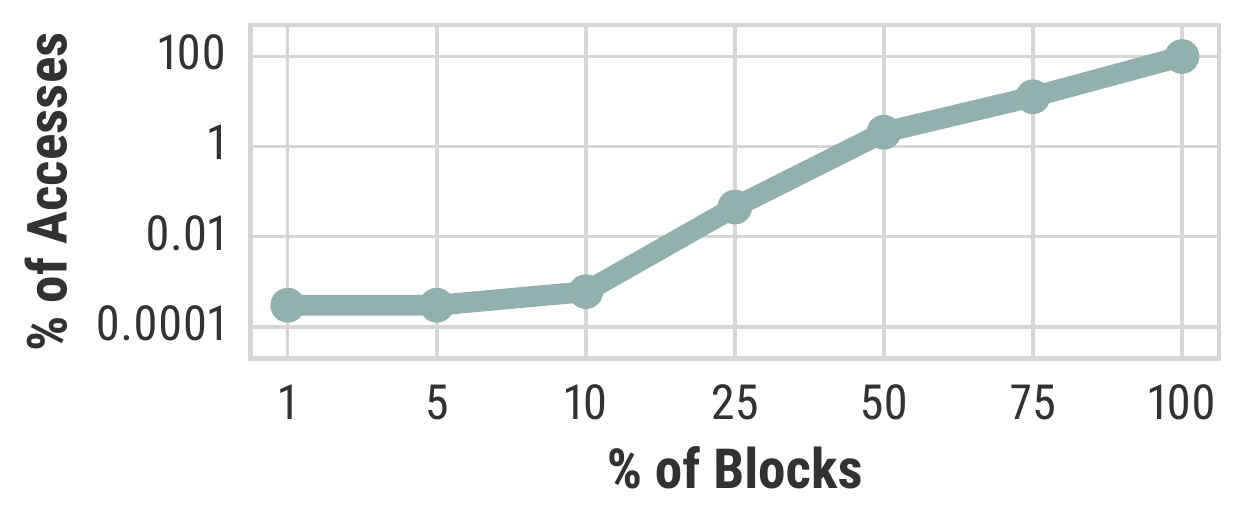}
        \label{fig:skew-tpcc}
    }
    \hfill
    \subfloat[VOTER]{
        \includegraphics[width=0.22\textwidth]
                        {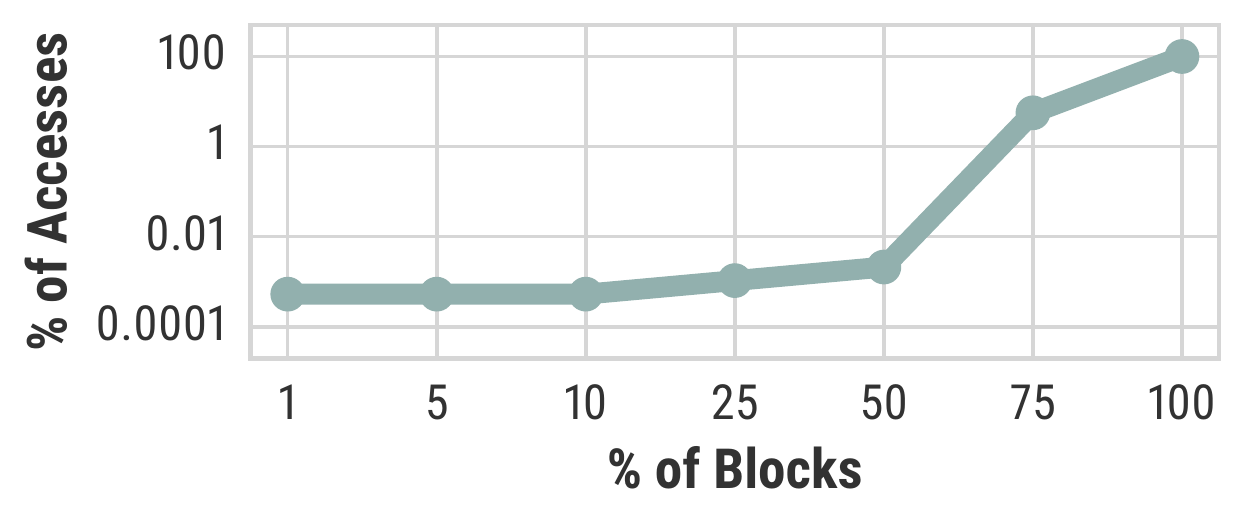}
        \label{fig:skew-voter}
    }                  
    \hfill
    \subfloat[AUCTIONMARK]{
        \includegraphics[width=0.22\textwidth]
                        {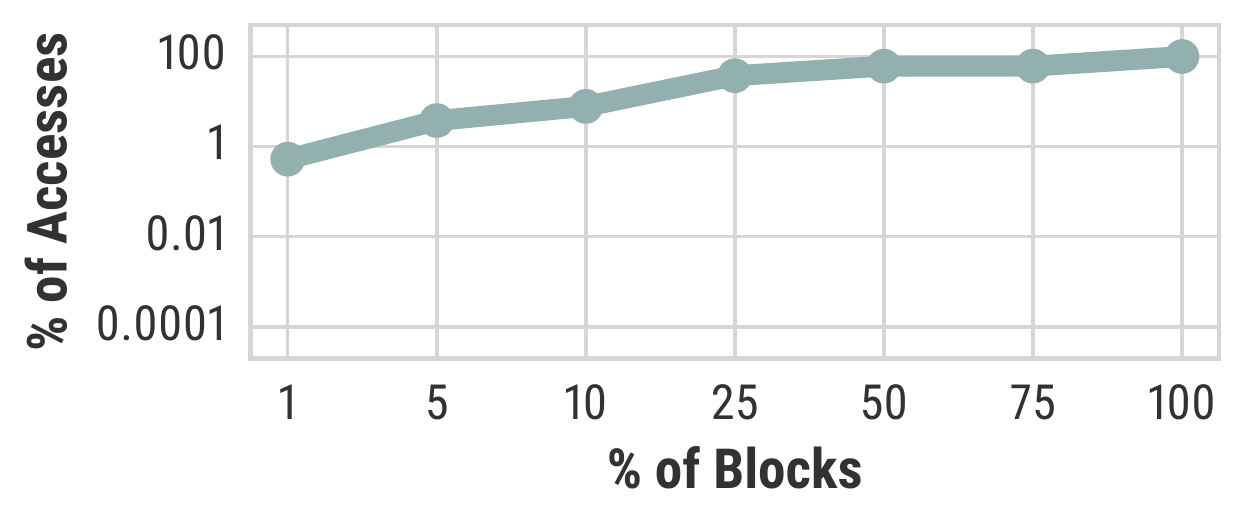}
        \label{fig:skew-auctionmark}
    }                         
   	\hfill
    \subfloat[CHBENCHMARK]{
        \includegraphics[width=0.22\textwidth]
                        {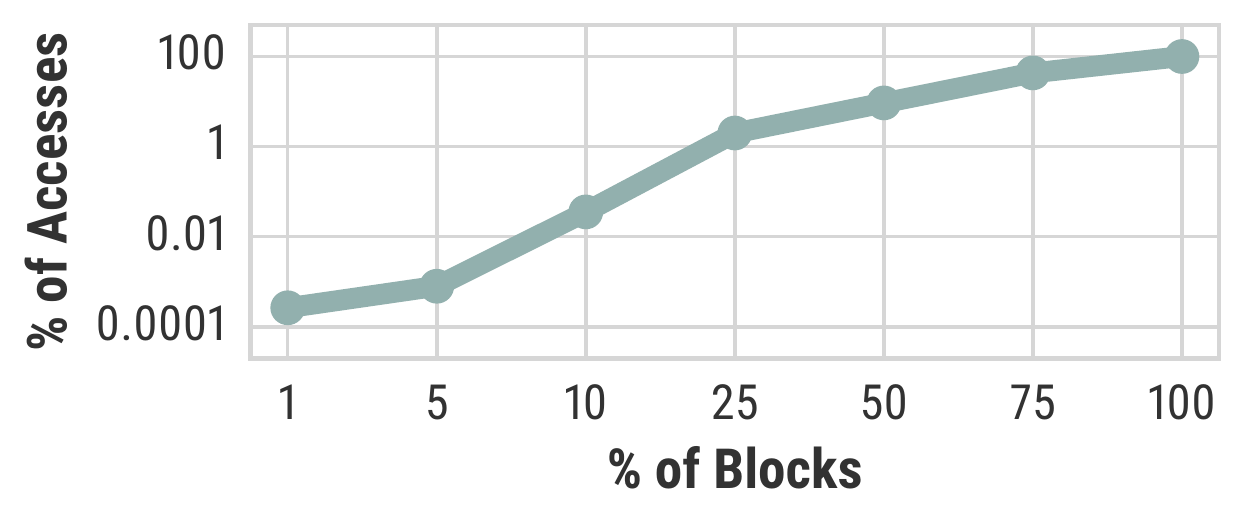}
        \label{fig:skew-chbenchmark}
    }
    \caption{
        \textbf{Workload Skew Characterisation:} CDF of the number of times a
        block is referenced in the traces associated with different workloads. Due to
        the inherent skew present in certain workloads, most of the buffer pool
        accesses are made to a small fraction of blocks.}
    \label{fig:skew}
\end{figure}

We begin with a characterization of the workload skew present in the different
workloads. \cref{fig:skew} shows the  cumulative distribution function (CDF) of
the number of buffer pool accesses per block in the workload traces.

For the \benchTPCC benchmark shown in~\cref{fig:skew-tpcc}, 13\% of buffer pool
references are made to 75\% of the blocks and 25\% of the blocks only account
for 0.05\% of the accesses. This illustrates that this workload is not highly
skewed and has a large working set. Similarly, the \benchChbenchmark also
exhibits low skew as depicted in~\cref{fig:skew-chbenchmark}. 50\% and 75\% of
the blocks account for 9\% and 43\%  of the buffer pool references,
respectively.

~\cref{fig:skew-voter} shows that the \benchVoter benchmark exhibits the lowest
degree of skew among all workloads since 75\% of the referenced blocks account
for only 6\% of buffer pool references. This is because the workload 
consists of short-lived transactions that generate writes to the log.
In contrast, \benchAuctionmark exhibits the highest degree of skew
among all workloads. 0.001\% of the blocks account for 8\% of the buffer pool
references and 61\% of the buffer pool accesses are made to 25\% of the blocks. 
We attribute this to the temporally skewed item access patterns in
\benchAuctionmark.
\\ \vspace{-0.05in}

\subsection{Impact of NVM on Runtime Performance}
\label{sec:runtime-performance}

In this section, we compare the buffer manager's throughput on similarly priced
NVM-SSD and DRAM-SSD storage hierarchies to examine the impact of NVM on 
runtime performance. We do not consider a DRAM-NVM-SSD hierarchy in this
experiment to isolate the utility of NVM. We configured the cost
budget to be \$10,000. Given this budget, the capacity of the NVM and DRAM
devices are 128~GB and 1~TB, respectively\footnote{The cost of NVM is derived
from the price of Intel's 3D XPoint-based Optane SSD 905P~\cite{optane18}}. 
The latter device's capacity is 8$\times$ higher than that of the
former due to NVM's higher capacity-cost ratio.
To obtain insights that are applicable for a wider range of NVM technologies,
we quantify the impact of NVM on different latency configurations. 
We ran the experiment under three NVM latency configurations for the emulator
ranging from 2--8$\times$ DRAM latency (320--1280~ns).

\begin{figure}[t!]
    \centering
    \fbox{\includegraphics[width=0.45\textwidth,trim=2.5cm 0.5cm 0.3cm 0.3cm]
            {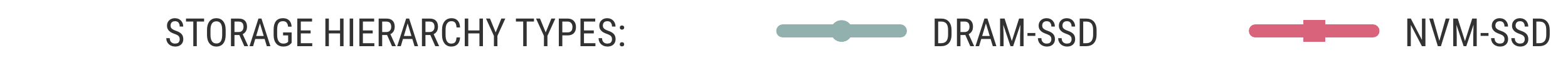}}
    \\[-0.1ex] 
    \subfloat[TPCC]{
        \includegraphics[width=0.22\textwidth]
                        {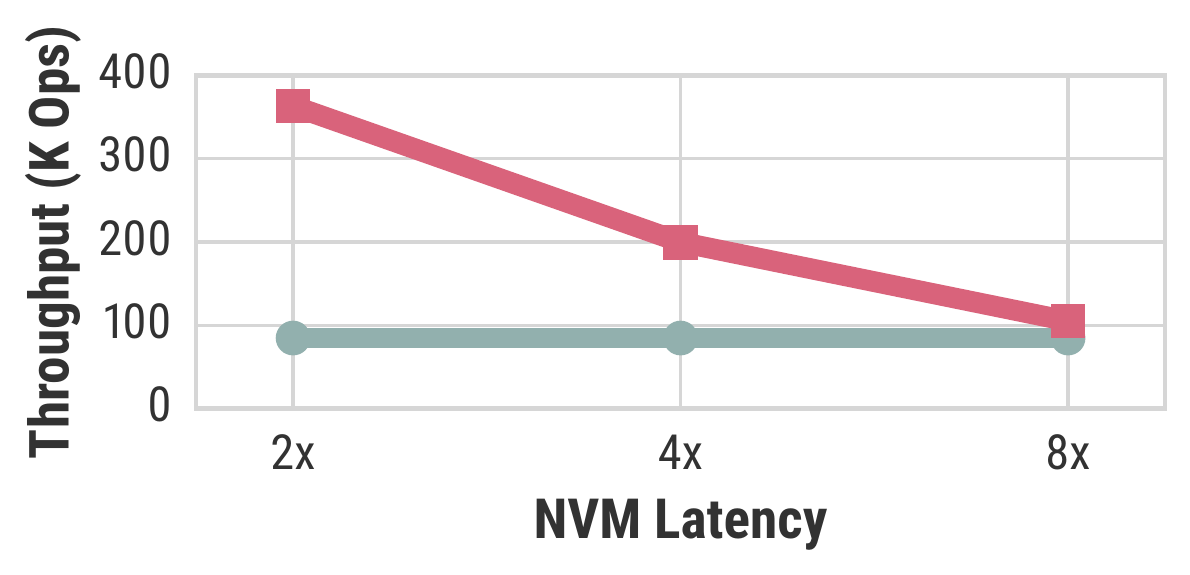}
        \label{fig:latency-tpcc}
    }    
   	\hfill
    \subfloat[VOTER]{
        \includegraphics[width=0.22\textwidth]
                        {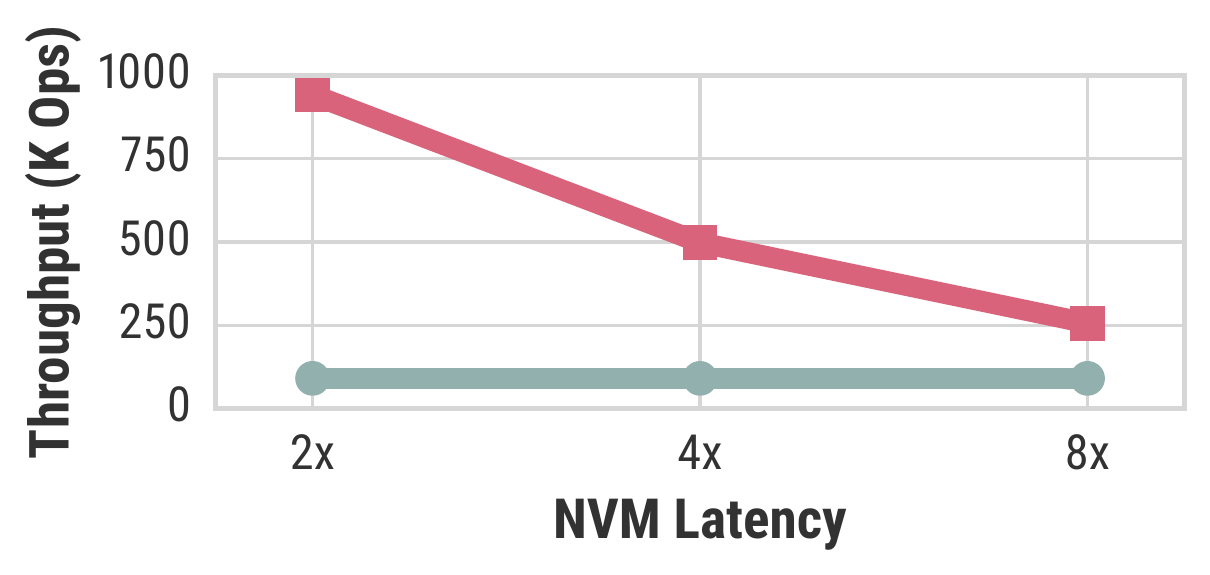}
        \label{fig:latency-voter}
    }                  
    \hfill
    \subfloat[AUCTIONMARK]{
        \includegraphics[width=0.22\textwidth]
                        {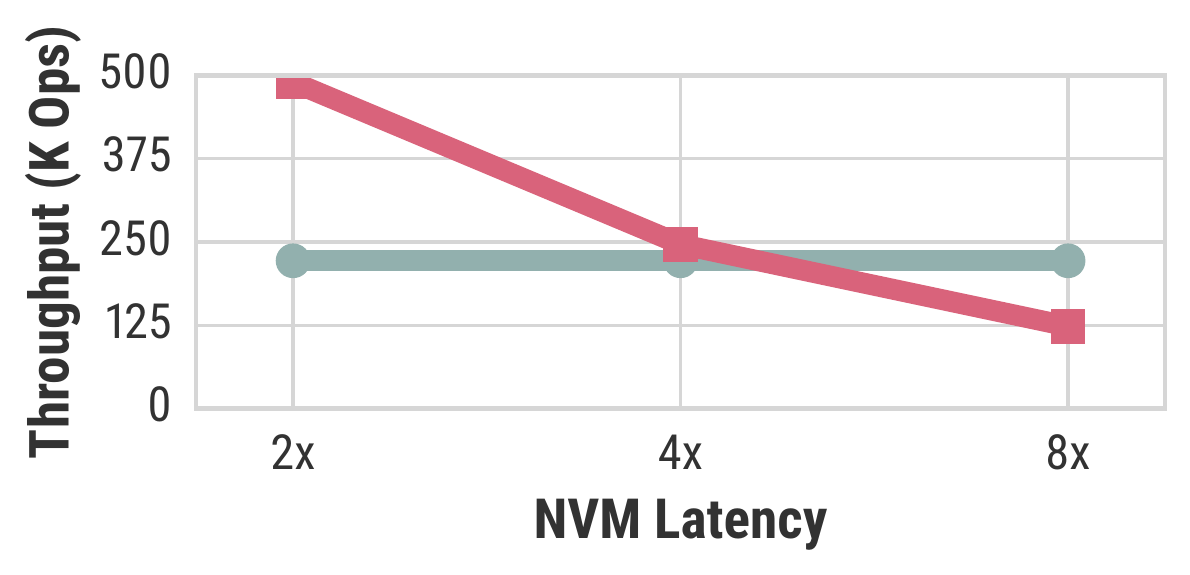}
        \label{fig:latency-auctionmark}
    }                         
    \hfill
    \subfloat[CHBENCHMARK]{
        \includegraphics[width=0.22\textwidth]
                        {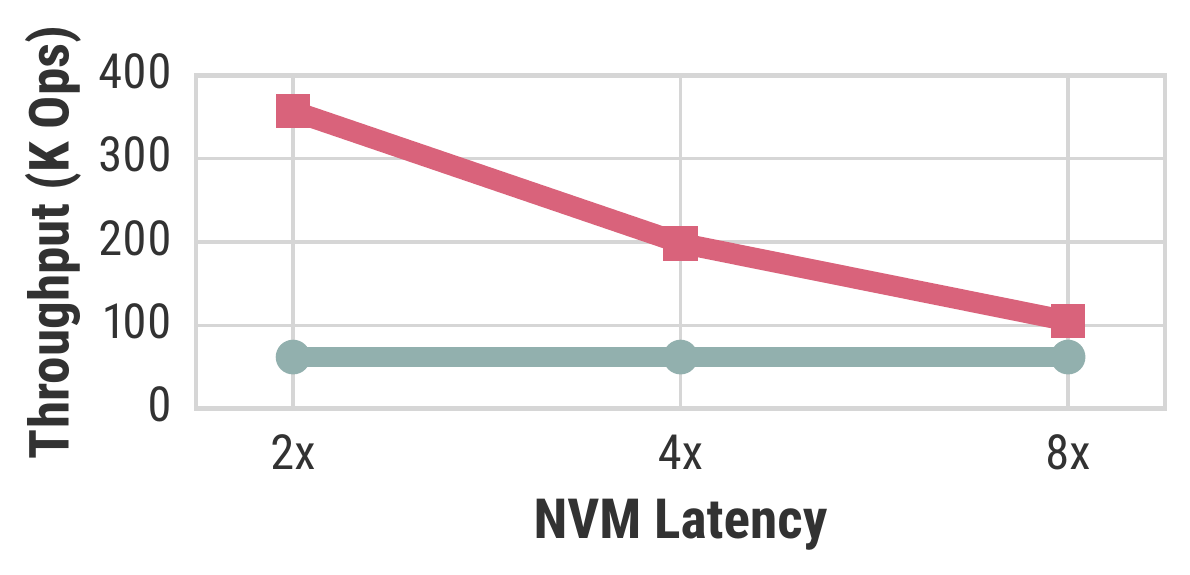}
        \label{fig:latency-chbenchmark}
    }
    \caption{
        \textbf{Impact of NVM on Runtime Performance:} Comparison of the
        buffer manager's throughput on similarly priced NVM-SSD and DRAM-SSD
        storage hierarchies under different NVM latency configurations.
    }
    \label{fig:latency}
\end{figure}

The results shown in~\cref{fig:latency} illustrate that the NVM-SSD hierarchy
outperforms its DRAM-based counterpart on most workloads and latency
configurations.
On the \benchTPCC benchmark, we observe that with the 2$\times$ latency
configuration, the NVM-based hierarchy outperforms the DRAM-SSD hierarchy by
4.3$\times$. This is because NVM reduces the number of SSD accesses 
by 19$\times$ due to its capacity advantage over DRAM. The reduction in time
spent on disk operations overrides the performance impact of slower NVM 
accesses. With the 4$\times$ latency configuration, the performance gap drops to
2.1$\times$. This illustrates the impact of NVM's higher latency relative to
DRAM. The 8$\times$ latency configuration is the break-even point at which 
both storage hierarchies deliver comparable throughput. In this setting,
slower NVM accesses nullify the benefits of its higher capacity.

The impact of NVM is more pronounced on the \benchVoter benchmark. 
This benchmark saturates the DBMS with short-lived transactions that each
update a small number of tuples. The buffer manager frequently flushes 
dirty blocks to durable storage while executing this workload. 
NVM improves runtime performance by efficiently absorbing these writes.
As shown in~\cref{fig:latency-voter}, the performance gap between the two
storage hierarchies varies from 10.5$\times$ to 2.8$\times$ on the 2$\times$ and
8$\times$ latency configurations, respectively.
   
On the \benchAuctionmark workload shown in~\cref{fig:latency-auctionmark},
the NVM-SSD hierarchy outperforms its DRAM-based counterpart by 2.2$\times$ with
the 2$\times$ latency configuration. However, the trend reverses on the
8$\times$ latency configuration, where the latter hierarchy delivers 1.8$\times$
higher throughput than the former. We attribute this to the workload's smaller
working set that fits in the DRAM buffer. So, the NVM buffer is not as
beneficial on this workload, particularly with slower latency configurations.

The results for the \benchChbenchmark workload, shown
in~\cref{fig:latency-chbenchmark}, illustrate that the NVM-based hierarchy
delivers 5.9$\times$ higher throughput compared to its DRAM-based counterpart on
the 2$\times$ latency configuration. We attribute this to the larger working set
associated with this workload. Even on the 8$\times$ latency configuration, 
the former storage hierarchy delivers 1.7$\times$ higher throughput than the
latter. This demonstrates the performance impact of NVM on HTAP workloads.

\subsection{Data Migration Policies}
\label{sec:exps-migration}

In this section, we look at the impact of data migration policies on runtime
performance and the number of writes performed on NVM. 
We begin by comparing the performance of the buffer manager when it employs the 
lazy and eager policies presented in~\cref{sec:buffer-management}.  
We consider a storage hierarchy with 16~GB DRAM and 1~TB NVM buffers on top of
SSD. We quantify the performance impact of four data flow optimizations: 
(1) bypassing DRAM ($\mathcal{D}_{r}$, $\mathcal{D}_{w}$), and (2) bypassing NVM
($\mathcal{N}_{r}$, $\mathcal{N}_{w}$) while serving read and write operations.
To derive insights that are applicable for a wider range of NVM technologies,
we do this analysis across three NVM latency configurations ranging from
2--8$\times$ DRAM latency.
\\ \vspace{-0.05in}

\begin{figure}[t!]
    \centering
    \subfloat[TPCC]{
        \includegraphics[width=0.21\textwidth]
                        {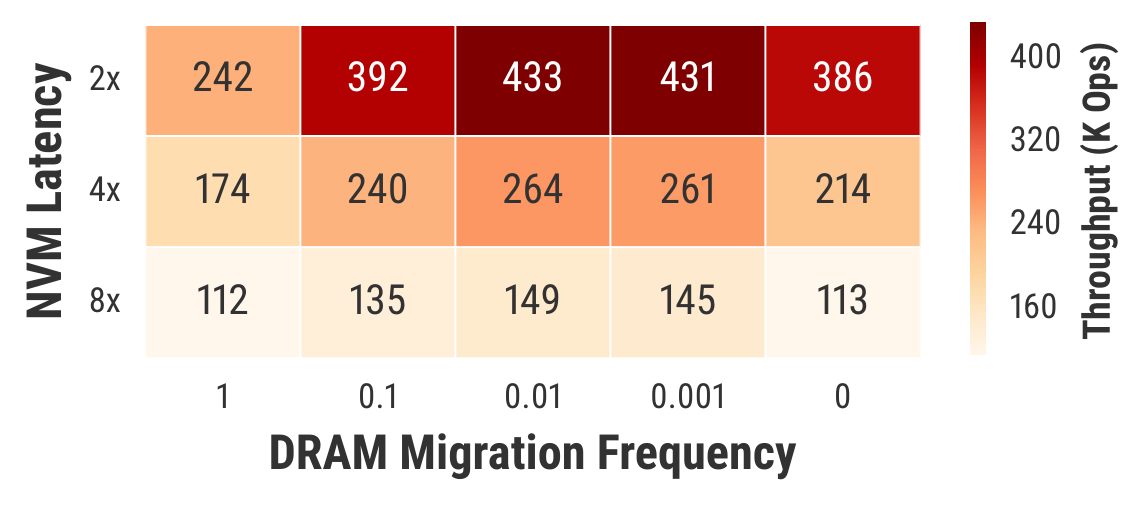}
        \label{fig:migration-dram-tpcc}
    }
    \hfill
    \subfloat[VOTER]{
        \includegraphics[width=0.22\textwidth]
                        {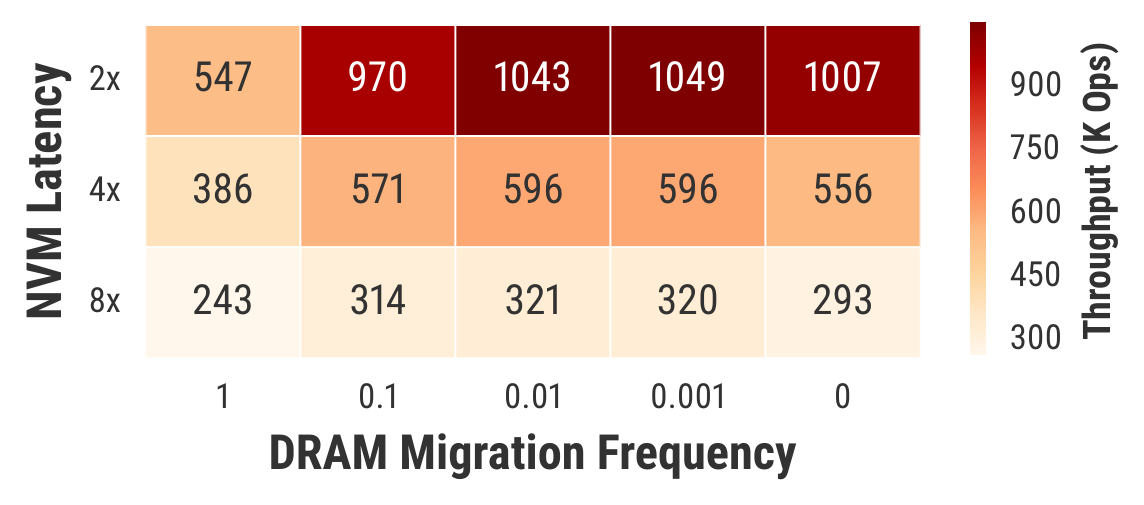}
        \label{fig:migration-dram-voter}
    }                  
    \hfill
    \subfloat[AUCTIONMARK]{
        \includegraphics[width=0.22\textwidth]
                        {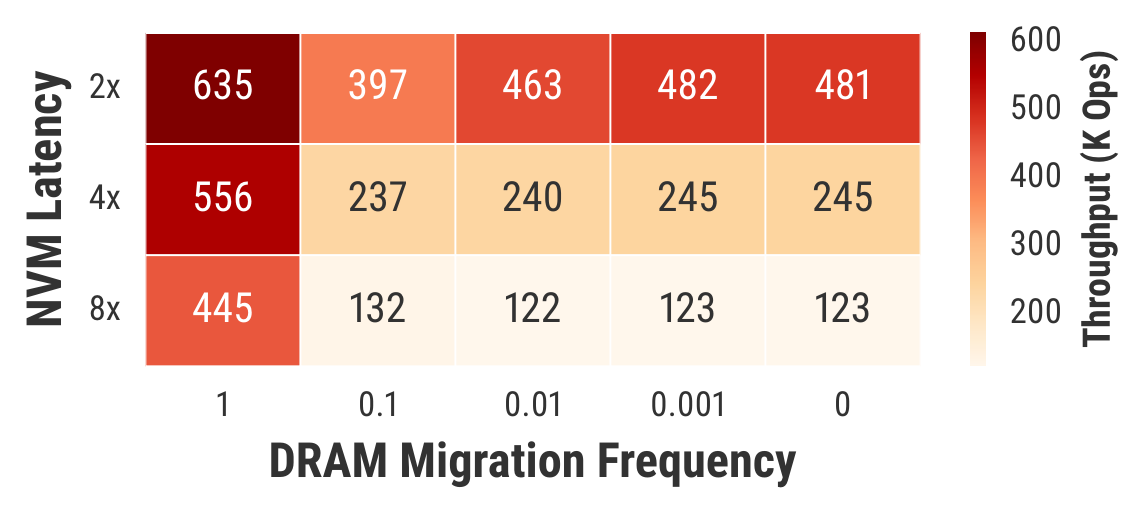}
        \label{fig:migration-dram-auctionmark}
    }            
   	\hfill
    \subfloat[CHBENCHMARK]{
        \includegraphics[width=0.22\textwidth]
                        {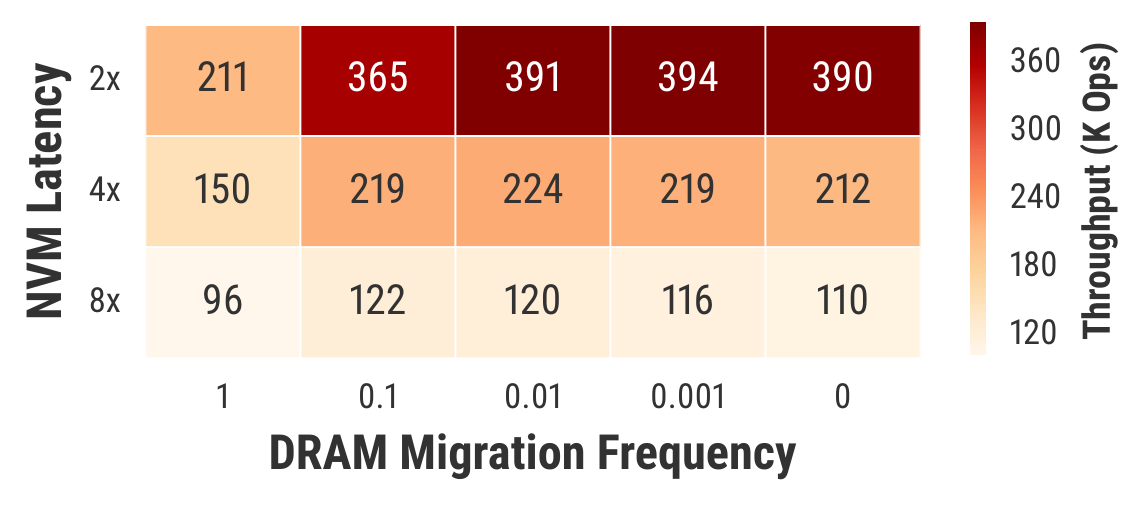}
        \label{fig:migration-dram-chbenchmark}
    }
    \caption{
        \textbf{Performance Impact of Bypassing DRAM:} 
		Comparison of the buffer manager's throughput when it adopts lazy and eager  
        data migration policies for DRAM. We measure the performance impact of
        these policies across different NVM latency configurations and DRAM
        migration frequencies ($\mathcal{D}$).
        }
    \label{fig:migration-dram}
\end{figure}

\textbf{Performance Impact of Bypassing DRAM}
~\cref{fig:migration-dram} illustrates the performance impact of bypassing DRAM
while serving reads and write operations. We vary the DRAM migration
probabilities ($\mathcal{D}_{r}$, $\mathcal{D}_{w}$) in lockstep from 1 through
0.
We configured the buffer manager to adopt an eager policy for NVM
($\mathcal{N}_{r}$, $\mathcal{N}_{w}$ = 1).
Since the DRAM migration probabilities are updated in lockstep in this
experiment, we denote them by $\mathcal{D}$. With the baseline policy ($\mathcal{D}$ = 1),
the buffer manager eagerly moves data to DRAM. The results
in~\cref{fig:migration-dram} demonstrate that the lazy migration policies work
well for DRAM on most workloads.

For the \benchTPCC benchmark  shown in~\cref{fig:migration-dram-tpcc}, the
throughput observed when $\mathcal{D}$ is 0.01 is 79\% higher than that
with the eager migration policy on the 2$\times$ latency configuration. 
The reasons for this are twofold.
First, the lazy policy reduces the data migration between NVM and DRAM. 
Second, it ensures that only frequently referenced data are moved to DRAM.
The performance gap drops to 33\% on the 8$\times$ latency configuration. 
This is because the lazy policy amplifies the performance impact of slower 
NVM operations.

The benefits of lazy data migration are more prominent on the
write-intensive \benchVoter workload. Bypassing DRAM while performing writes
nearly doubles the throughput, as shown in~\cref{fig:migration-dram-voter}.
With the lazy policy, the buffer manager directly flushes dirty blocks to NVM
instead of first writing them on DRAM. Since DRAM write latencies are comparable
to those of NVM, particularly on the 2$\times$ latency configuration, bypassing
DRAM during writes shrinks the overall write latency.

Unlike other workloads, eager policy works well for the \benchAuctionmark
workload, as depicted in ~\cref{fig:migration-dram-auctionmark}. 
It outperforms the lazy policy ($\mathcal{D}$ = 0.1) by 60\% on the 2$\times$
latency configuration. This is because the workload's working set fits in the
DRAM buffer and shifts over time. But, the lazy policy delays the migration of
hot data from NVM to DRAM, thereby reducing the utility of the DRAM buffer. 
The performance gap shrinks to 32\% with a lazier policy ($\mathcal{D}$ =
0.001). The reduction in data movement between DRAM and NVM dampens the impact
of delayed migration of the working set.

Lastly, on the \benchChbenchmark workload, lazy policy delivers 85\% higher
throughput than its eager counterpart, as shown
in~\cref{fig:migration-dram-chbenchmark}. The working set of this workload is
comparatively more stable. So, even though the lazy policy results in delayed
migration, the buffer manager eventually loads the working set in the DRAM 
buffer. Thus, the optimal migration policy depends on the workload
characteristics.
\\ \vspace{-0.05in}

\begin{figure}[t!]
    \centering
    \subfloat[TPCC]{
        \includegraphics[width=0.21\textwidth]
                        {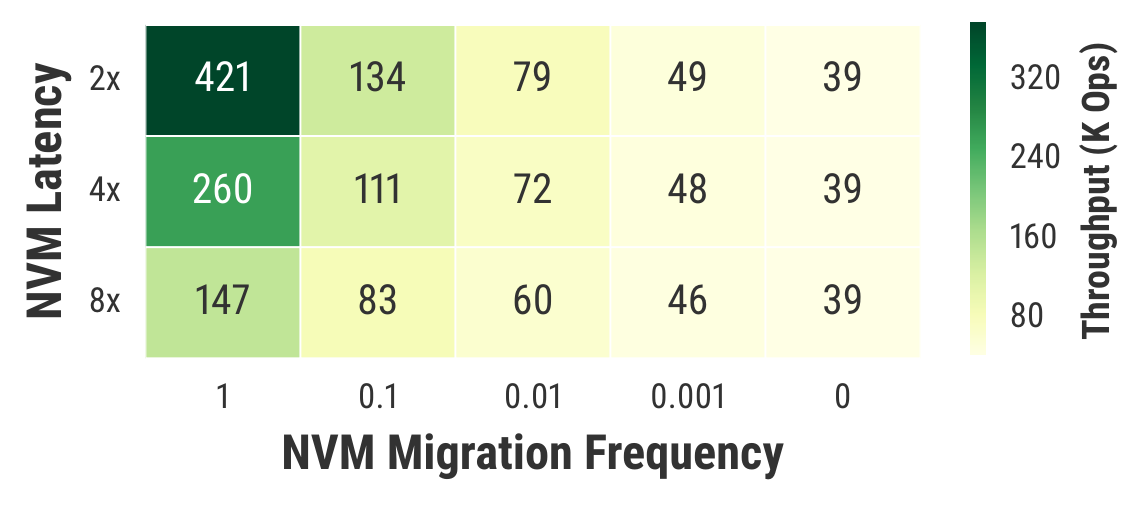}
        \label{fig:migration-nvm-tpcc}
    }
    \hfill
    \subfloat[VOTER]{
        \includegraphics[width=0.22\textwidth]
                        {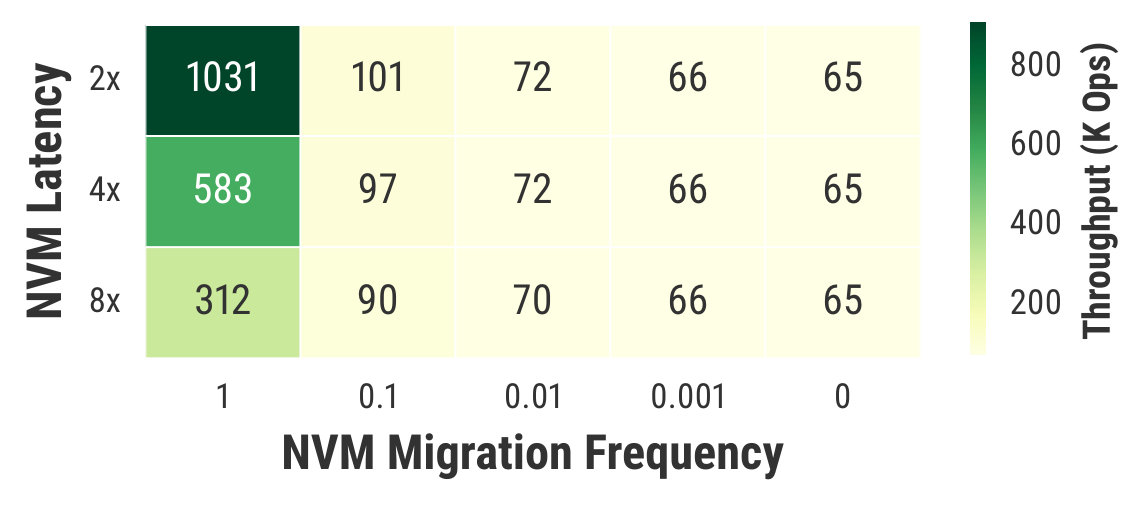}
        \label{fig:migration-nvm-voter}
    }                  
    \caption{
        \textbf{Performance Impact of Bypassing NVM:} 
        Comparison of the buffer manager's throughput when it adopts lazy and eager  
        data migration policies for NVM. We measure the performance impact of
        these policies across different NVM latency configurations and NVM
        migration frequencies ($\mathcal{N}$).
        }
    \label{fig:migration-nvm}
\end{figure}

\textbf{Performance Impact of Bypassing NVM}
~\cref{fig:migration-nvm} illustrates the performance impact of bypassing NVM
while serving reads and write operations.
In this experiment, we vary the NVM migration probabilities ($\mathcal{N}_{r}$,
$\mathcal{N}_{w}$) in lockstep from 1 through 0. We configured the buffer
manager to adopt an eager policy for DRAM ($\mathcal{D}_{r}$, $\mathcal{D}_{w}$
= 1). Since the NVM migration probabilities are updated in lockstep, we denote
them by $\mathcal{N}$. The results in~\cref{fig:migration-nvm} show that eager
migration ($\mathcal{N}$ = 1) works well for NVM on most workloads.

For the \benchTPCC benchmark shown in~\cref{fig:migration-nvm-tpcc}, 
the throughput observed when $\mathcal{N}$ is set to 0.1 is 68\% lower than that
with the eager policy on the 2$\times$ latency configuration. 
This is because the time spent on SSD operations increases by 15$\times$ 
due to bypassing NVM during writes. The performance impact of lazy migration
marginally drops to 43\% on the 8$\times$ latency configuration. Slower NVM
accesses dampen the effect of writes landing on SSD with this configuration. 

The performance impact of NVM bypass is more prominent on the \benchVoter
workload shown in~\cref{fig:migration-nvm-voter}. The throughput drops by 91\%
when $\mathcal{N}$ is set to 0.1 on the 2$\times$ latency configuration.
These results illustrate that while lazy migration policies work well for DRAM,
eager policies are a better fit for NVM.
\\ \vspace{-0.05in}

\begin{figure}[t!]
    \centering
    \subfloat[TPCC]{
        \includegraphics[width=0.21\textwidth]
                        {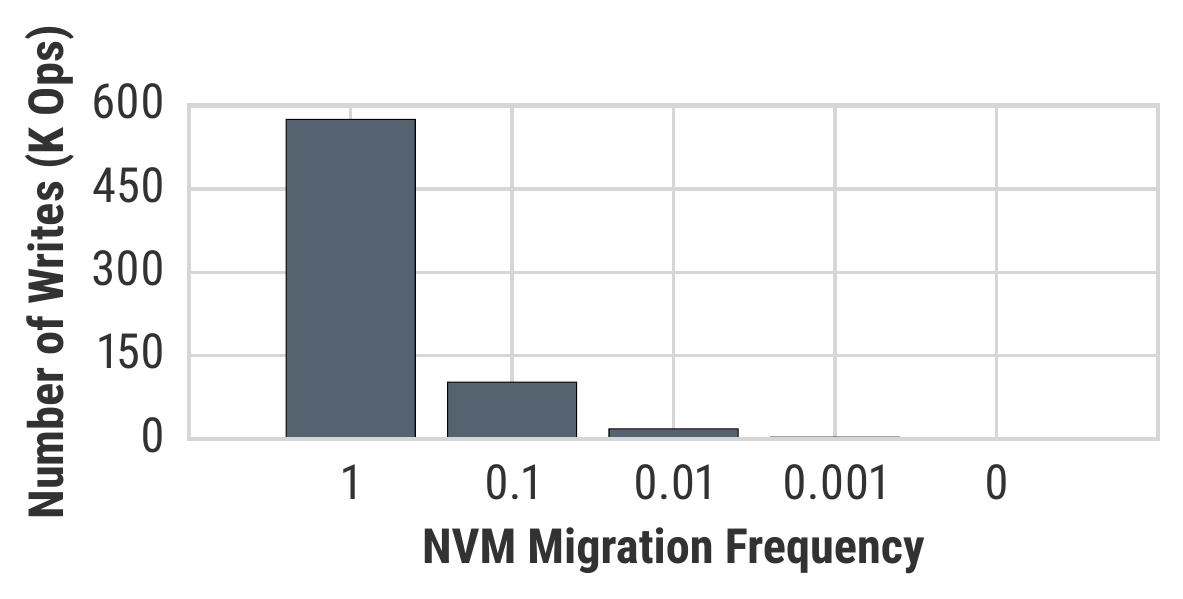}
        \label{fig:write-tpcc}
    }
   	\hfill
    \subfloat[VOTER]{
        \includegraphics[width=0.22\textwidth]
                        {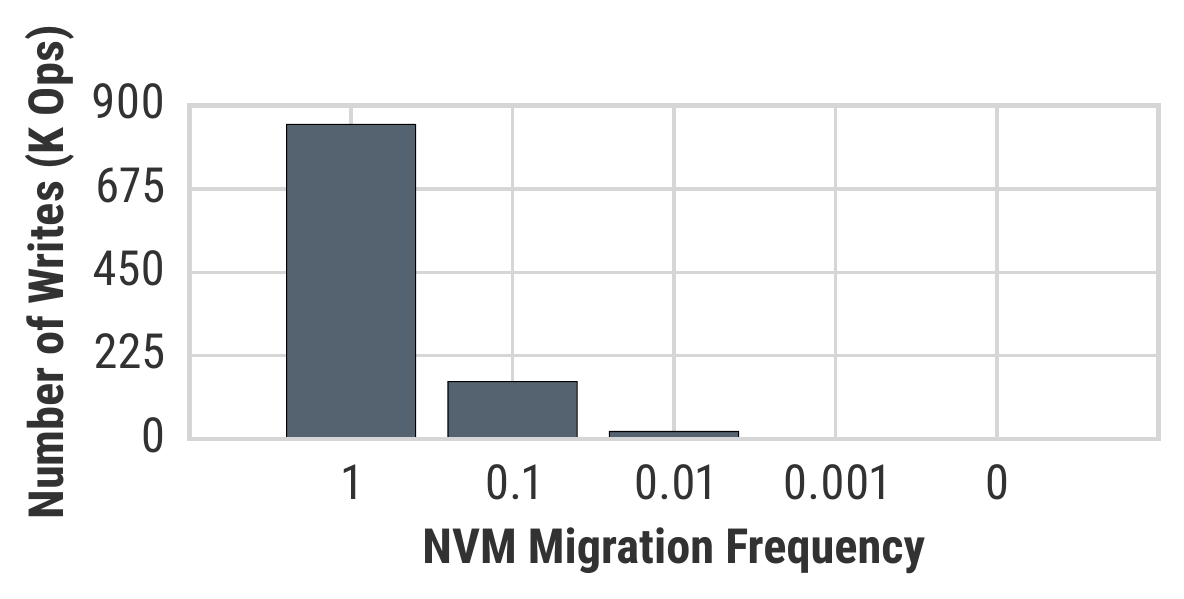}
        \label{fig:write-voter}
    }
    \caption{
        \textbf{Impact of Bypassing NVM on Writes to NVM:}         
        Comparison of the number of writes performed on NVM when the buffer
        manager adopts lazy and eager data migration policies for NVM. 
        We measure the impact of these policies across different 
        NVM migration frequencies ($\mathcal{N}$).
        }
    \label{fig:write}
\end{figure}

\textbf{Impact of NVM Bypass on Writes to NVM}
Although lazy data migration negatively impacts runtime performance, it 
reduces the number of writes performed on NVM.~\cref{fig:write} presents the
impact of NVM bypass on the number of NVM writes. 
For the \benchTPCC benchmark, as shown in~\cref{fig:write-tpcc}, the buffer
manager performs 5.5$\times$ fewer writes to NVM with a lazy migration policy
($\mathcal{N}$ = 0.1) in comparison to eager migration.
The impact of NVM bypass on the number of writes performed on NVM is equally  
pronounced on the \benchVoter workload as shown in~\cref{fig:write-voter}.
Adopting a lazy migration policy ($\mathcal{N}$ = 0.1) reduces the number of NVM
writes by 8.5$\times$.

These results illustrate that the optimal data migration policy must be chosen
depending on the runtime performance requirements and write endurance
characteristics of NVM.
\\ \vspace{-0.05in}

\begin{figure}[t!]
    \centering
    \subfloat[TPCC (4~GB DRAM)]{
        \includegraphics[width=0.21\textwidth]
                        {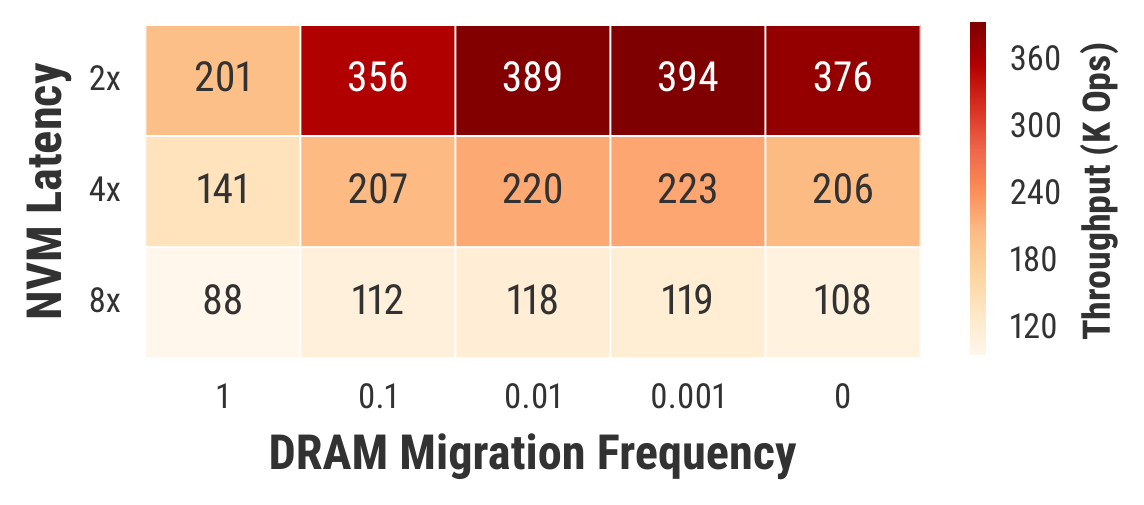}
        \label{fig:migration-small}
    }    
   	\hfill
    \subfloat[TPCC (64~GB DRAM)]{
        \includegraphics[width=0.21\textwidth]
                        {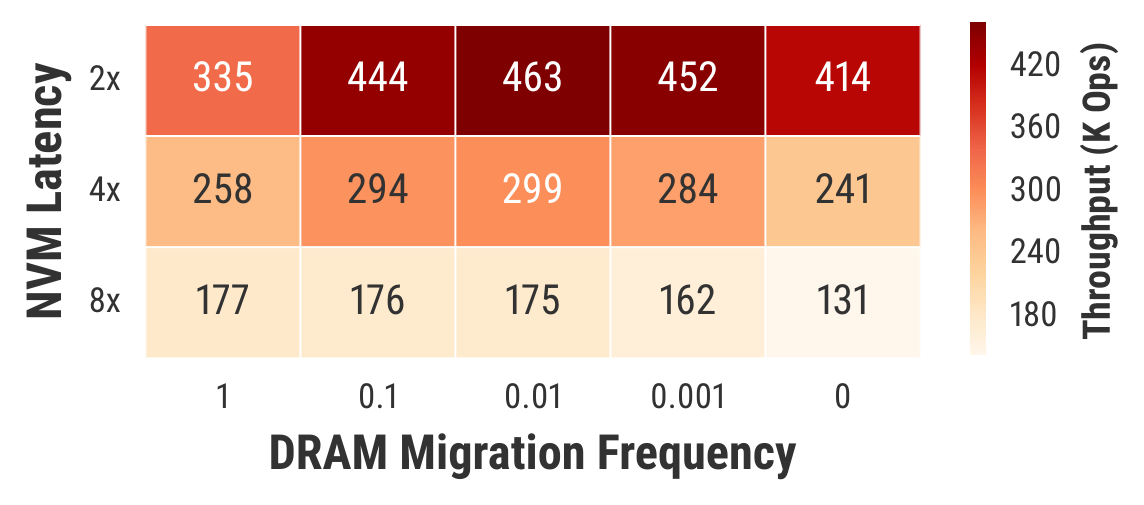}
        \label{fig:migration-large}
    }
    \caption{
        \textbf{Impact of Storage Hierarchy:} 
		Comparison of the optimal data migration policy decision for bypassing DRAM 
		across different storage hierarchies, NVM latency configurations, and DRAM
        migration frequencies.
        }
    \label{fig:migration-storage-hierarchy}
\end{figure}

\textbf{Impact of Storage Hierarchy}
We next consider how the optimal data migration policy varies across storage
hierarchies. In this experiment, we consider two three-tier storage
hierarchies with 4~GB and 64~GB DRAM buffers. We configured both systems 
to use a 1~TB NVM buffer on top of SSD. 
The results for the \benchTPCC benchmark depicted
in~\cref{fig:migration-storage-hierarchy} show that the utility of lazy data
migration varies across storage systems.

On the first system, as shown in~\cref{fig:migration-small}, the throughput
with lazy migration ($\mathcal{D}$ = 0.01) is 94\% higher than that with its
eager counterpart. The performance gap between the policies on this system is
larger than that observed with a larger DRAM buffer (16~GB)
in~\cref{fig:migration-dram-tpcc}. This is because the lazy policy
increases the utility of the smaller DRAM buffer by not polluting it with colder data. 
For this system, the optimal migration frequency remains unchanged even on
slower latency configurations.

The results for the second system shown in~\cref{fig:migration-large} illustrate
that the lazy policy delivers 38\% higher throughput on the 2$\times$ latency
configuration. The utility of lazy migration is not as prominent on this system
since the capacity of the DRAM buffer is one-sixteenth of that of the NVM
buffer. The performance with the eager policy ($\mathcal{D}$ = 1) is comparable
to that with its lazy counterpart on the 8$\times$ latency configuration.
This is because the latter policy amplifies the impact of slower NVM operations, 
particularly when the relative size of the DRAM buffer compared to the NVM
buffer is large.

These results show that the optimal migration policy depends not only on the
workload and device characteristics, but also on the relative size of the DRAM
buffer compared to the NVM buffer.

\subsection{Adaptive Data Migration}
\label{sec:exps-tuning}

\begin{figure}[t!]
    \centering
    \fbox{\includegraphics[width=0.43\textwidth,trim=2.5cm 0.5cm 0.3cm 0.3cm]
            {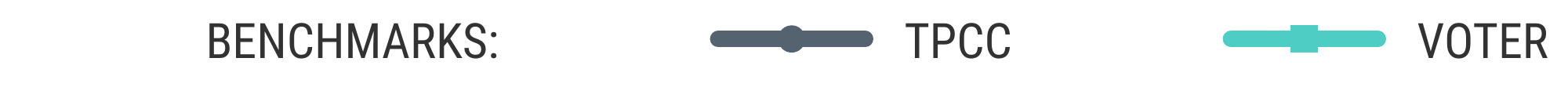}}
    \\[1ex] 
    \includegraphics[width=0.47\textwidth]
                    {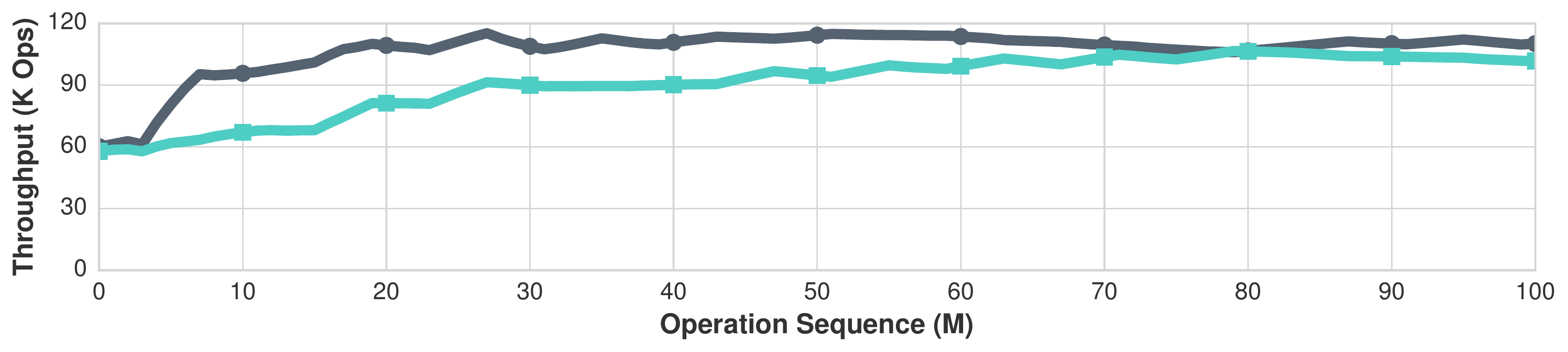}
    \label{fig:sync-tpcc}
    \caption{
		\textbf{Adaptive Data Migration:} The impact of buffer management policy
		adaptation on runtime performance across different workloads.
    }    
    \label{fig:tuning}
\end{figure}

In the previous experiments, we examined the utility of a \textit{fixed} data
migration policy. In the real world, identifying the optimal data
migration policy is challenging due to diversity  of workloads and storage
hierarchies. Thus, we now examine the ability of buffer manager to
automatically adapt the management policy at runtime.
In this experiment, the buffer manager begins executing the workload with an
eager policy for both DRAM ($\mathcal{D}$ = 1) and NVM ($\mathcal{N}$ = 1).
During execution, it adapts the policy using the simulated annealing (SA)
algorithm presented in~\cref{sec:adaptive-migration}. 
This technique searches for the policy that maximizes the throughput given a
target workload and storage hierarchy. We use an operation sequence with 100~M
entries. We set $\alpha$ and $\gamma$ to 0.9 and 10, respectively. We
configure the initial and final temperatures of the annealing process to 800
and 0.00008. We configure  the duration of a tuning step to be 1~M operations to
ensure that the impact of policy changes are prominently visible to the SA
algorithm.

The results in~\cref{fig:tuning} show that the buffer manager converges to a
near-optimal policy for different workloads without requiring any manual tuning.
For the \benchTPCC and \benchVoter workloads, tuning the data migration policy
increases throughput by 79\% and 92\%, respectively. 
The buffer manager converges to a hybrid policy, with lazy migration for 
DRAM and eager migration for NVM on both workloads. The throughput converges to
a global optima over time. We attribute this to the gradual cooling mechanism in
SA that decreases the probability of accepting worse policies.

\subsection{Storage System Recommendation}
\label{sec:exps-size}

We next focus on the storage hierarchy recommendation problem presented 
in~\cref{sec:hierarchy-selection}. In this experiment, we compare the
\textit{performance/price numbers} of multi-tier storage hierarchies.
If the cost of a storage hierarchy is \$ $\mathcal{C}$ and the
throughput it delivers is $\mathcal{T}$ operations per second, then the
performance/price number is given by $\frac{\mathcal{T}}{\mathcal{C}}$. This
represents the number of operations executed per second per dollar. 
Given a system cost budget and a target workload, the recommender system
identifies the storage hierarchy with the highest performance/price number.

Each storage system consists of at most three devices: DRAM, NVM, and SSD. 
We vary the capacity of the DRAM and NVM devices from 4~GB through
64~GB, and from 512~GB through 2~TB, respectively. We configured the capacity
of the SSD device to 2~TB. We examine the runtime performance of the buffer
manager on both two- and three-tier storage hierarchies: DRAM-SSD, NVM-SSD, 
and DRAM-NVM-SSD. We configured the NVM latency to be 2$\times$ that of DRAM.
\\ \vspace{-0.05in}

\begin{figure}[t!]
    \centering
    \subfloat[STORAGE SYSTEM COST]{
	    \includegraphics[width=0.22\textwidth]
                {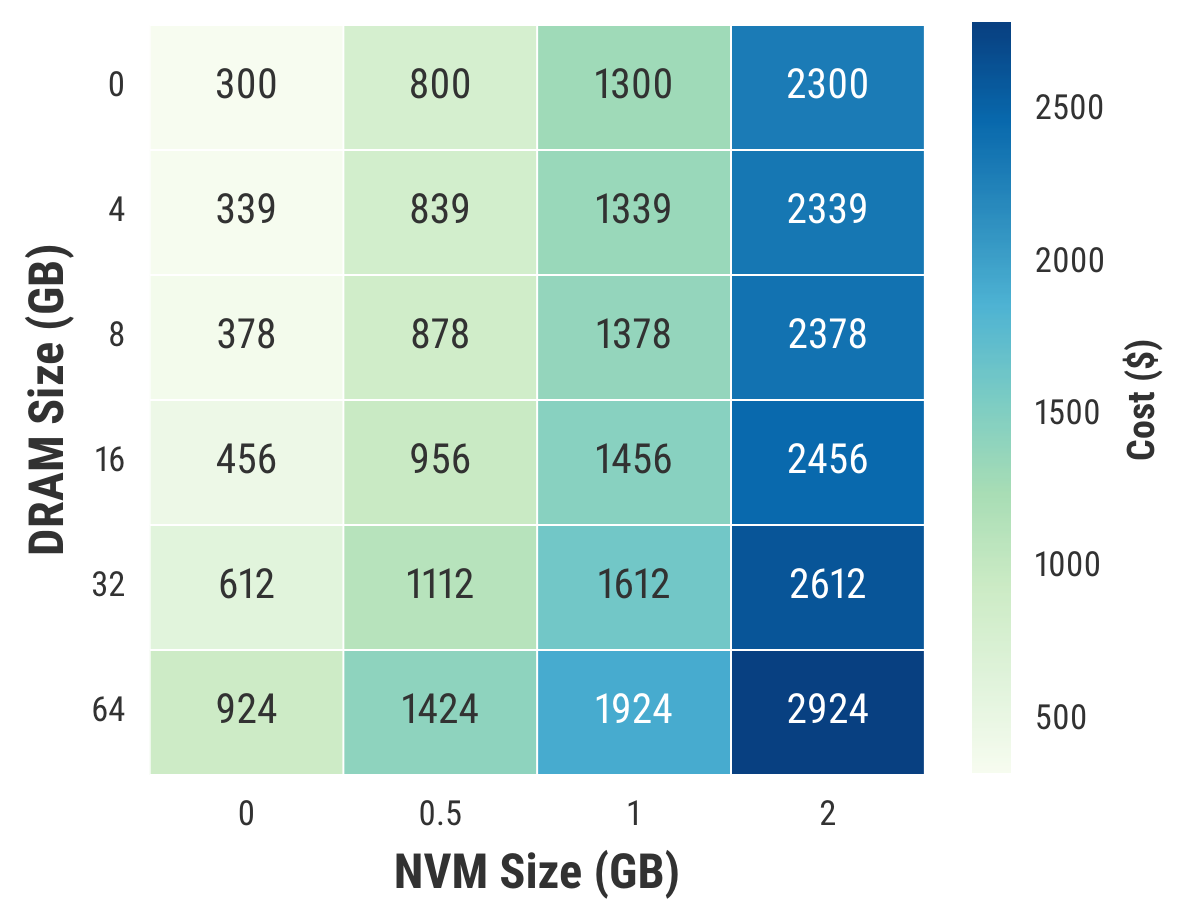}
			    \label{fig:cost}
	}
    \hfill
    \subfloat[TPCC]{
        \includegraphics[width=0.22\textwidth]
                        {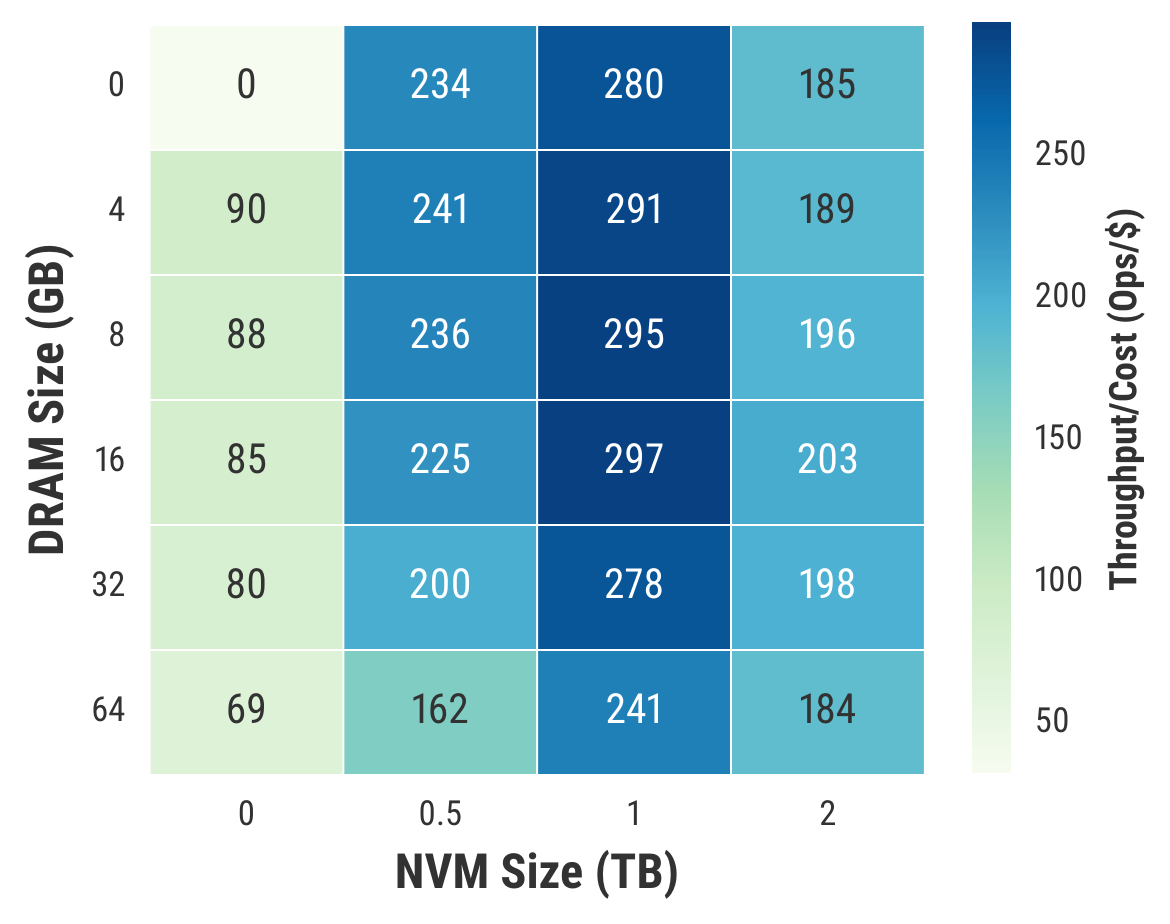}
        \label{fig:price-tpcc}
    }
    \hfill
    \subfloat[VOTER]{
        \includegraphics[width=0.22\textwidth]
                        {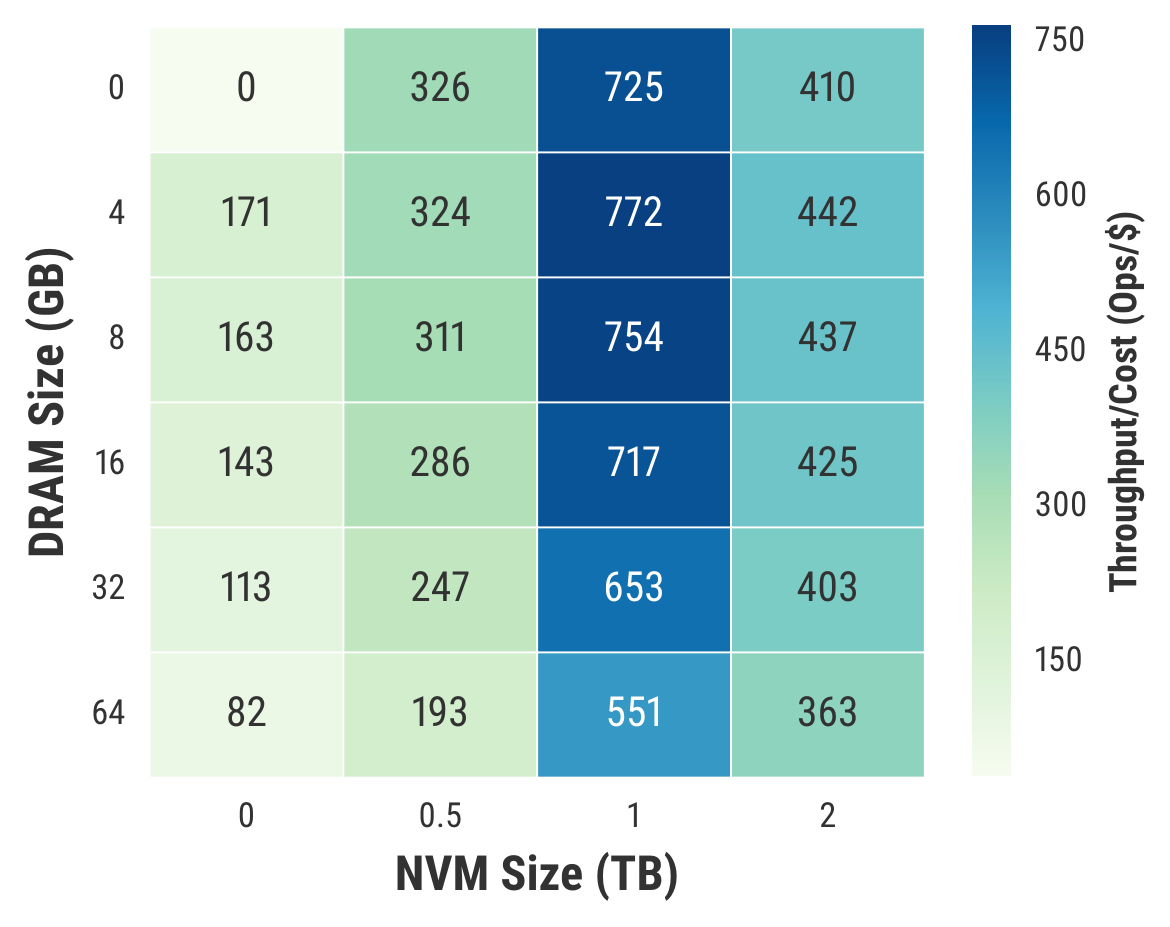}
        \label{fig:price-voter}
    }                  
    \hfill
    \subfloat[AUCTIONMARK]{
        \includegraphics[width=0.22\textwidth]
                        {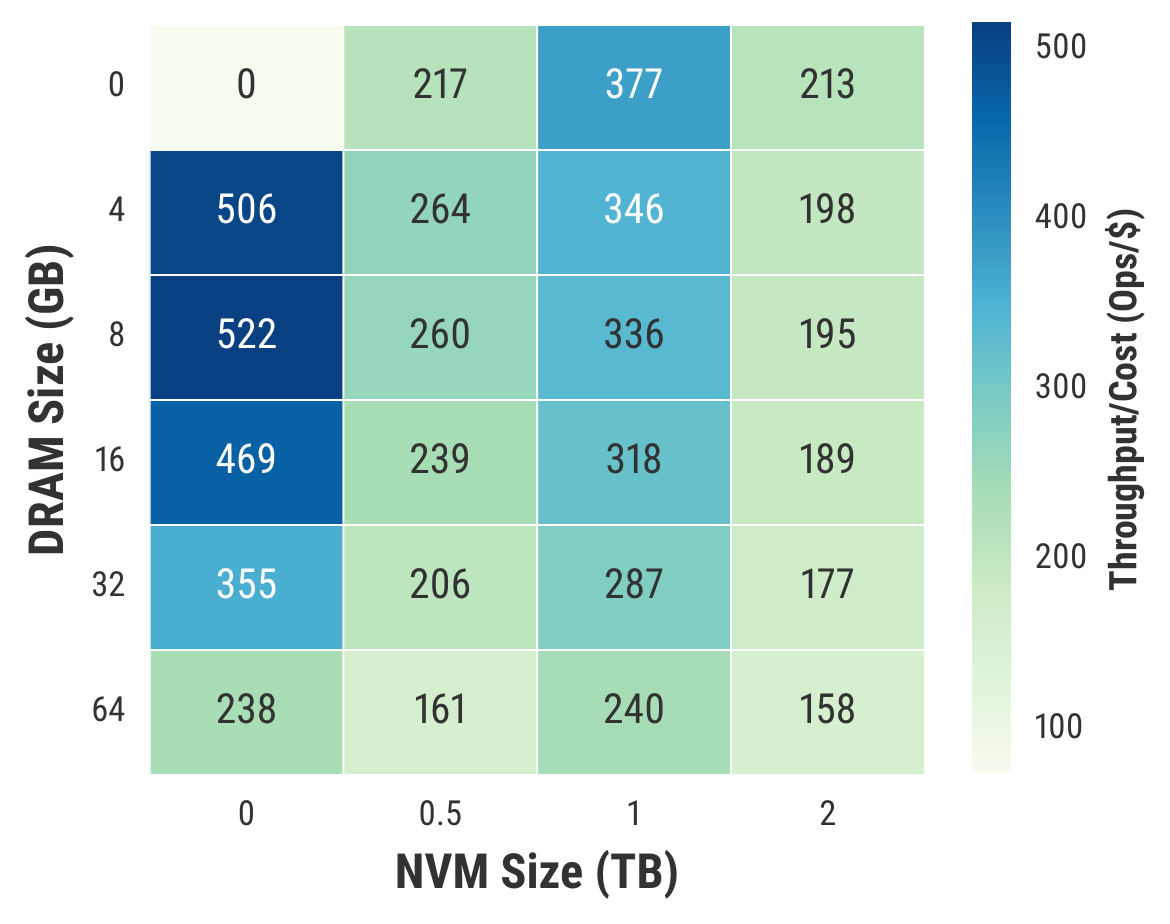}
        \label{fig:price-auctionmark}
    }                         
   	\hfill
    \subfloat[CHBENCHMARK]{
        \includegraphics[width=0.22\textwidth]
                        {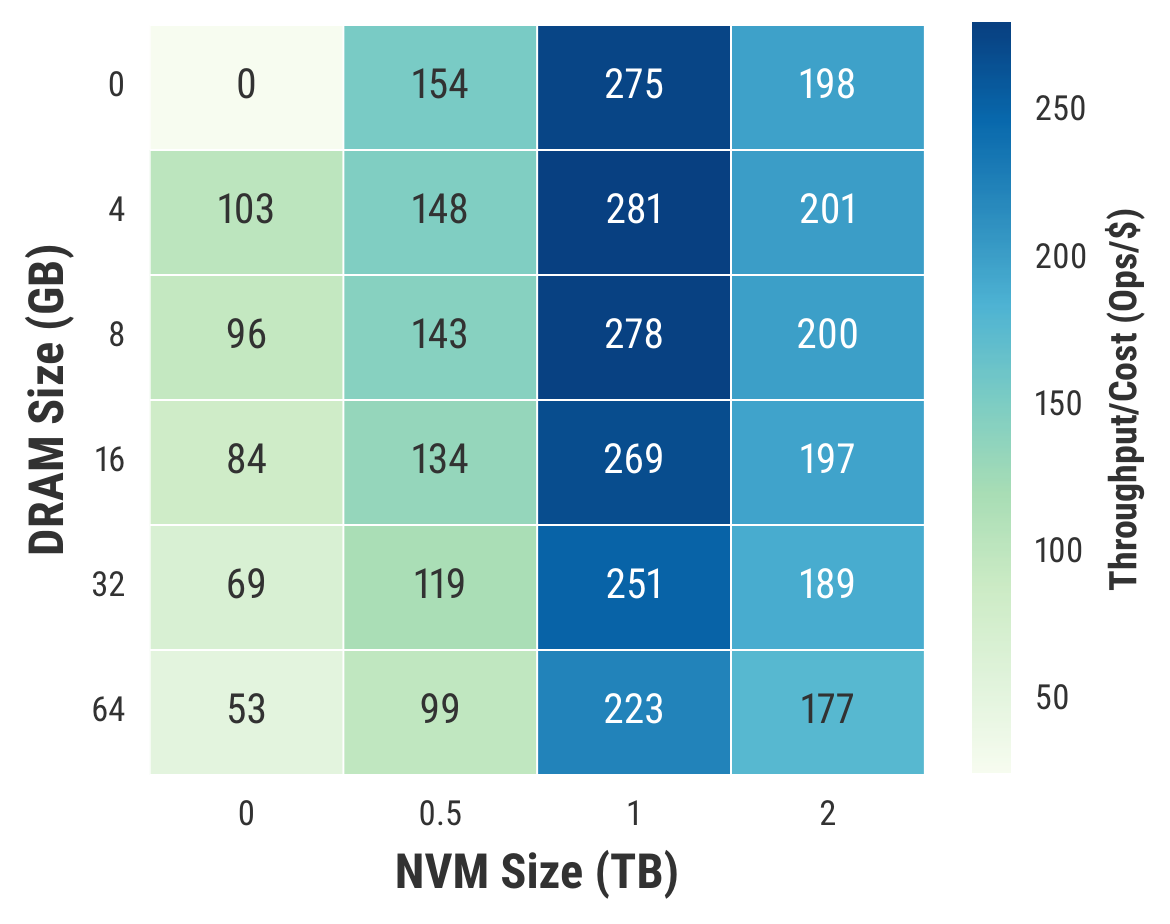}
        \label{fig:price-chbenchmark}
    }
   	\hfill
    \subfloat[TPCC (8$\times$ Latency)]{
        \includegraphics[width=0.22\textwidth]
                        {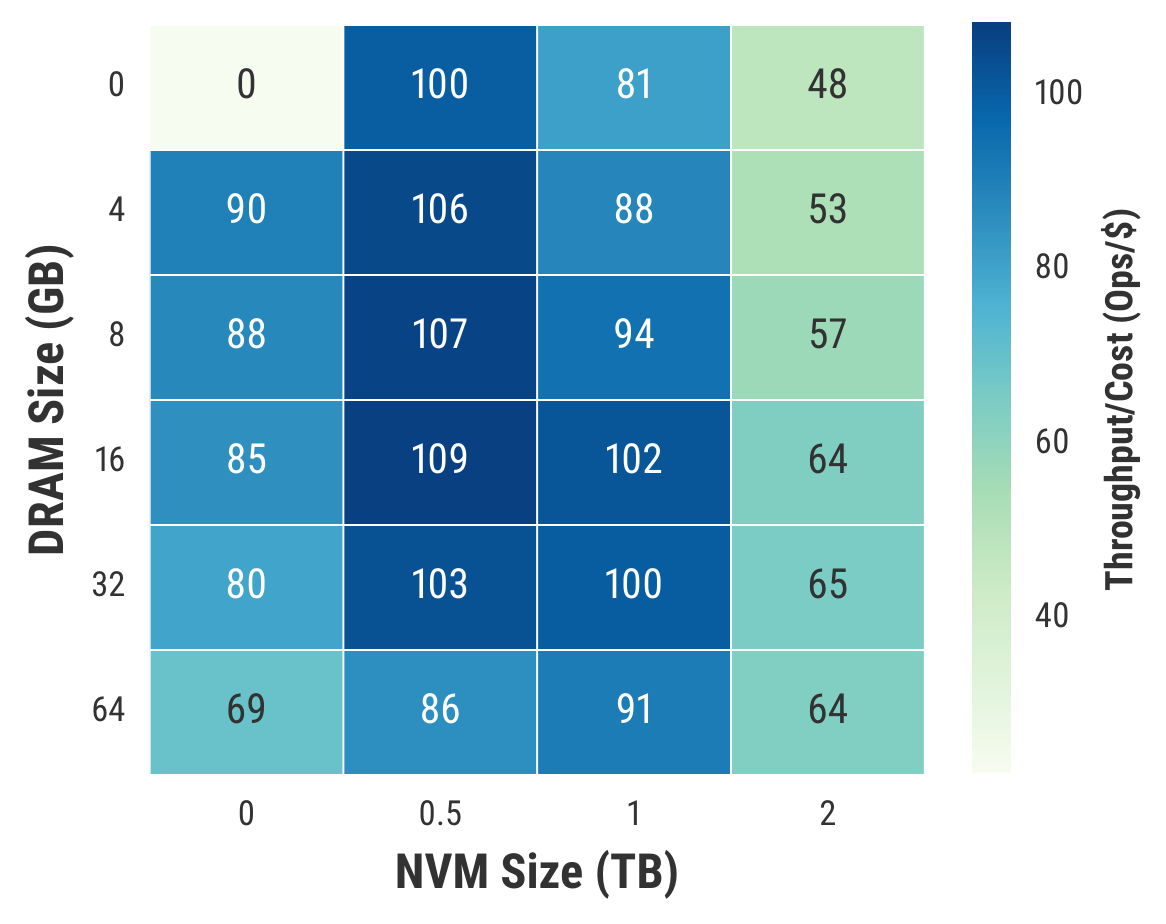}
        \label{fig:price-tpcc-8}
    }    
    \caption{
        \textbf{Storage Hierarchy Recommendation:} (a) The total cost of the
        DRAM, NVM, and SSD devices used in a multi-tier storage system. 
        (b-f) The performance/price numbers of candidate storage hierarchies on
        different benchmarks. Given a system cost budget and a target workload,
        the recommendation system performs a grid search to identify the storage 
        hierarchy with the highest performance/price number.
        }
    \label{fig:price}
\end{figure}

\textbf{Storage System Cost:} 
\cref{fig:cost} presents the cost of candidate storage hierarchies. 
The cost of the DRAM-SSD hierarchy raises from \$339 to \$924 when we
vary the capacity of the DRAM device from 2~GB through 32~GB. 
The cost of the NVM-SSD hierarchy increases from \$800 to \$2300 when we 
vary the capacity of the NVM device from 512~GB through 2~TB.
\\ \vspace{-0.05in}

\textbf{Storage Hierarchy Recommendation:} 
~\cref{fig:price} shows the performance/price numbers of candidate storage
hierarchies across different workloads. The recommender system performs a grid
search to identify the storage hierarchy with the highest performance/price
number on a target workload given a cost budget.

For the \benchTPCC benchmark, as shown in~\cref{fig:price-tpcc}, 
the storage system that delivers the highest performance/price number
consists of 16~GB DRAM and 1~TB NVM on top of SSD.
Expanding the capacity of the DRAM buffer to 64~GB improves performance by 7\%.
But, this also raises the storage system cost by 32\%.
Similarly, reducing the capacity of the DRAM buffer to 4~GB shrinks
performance and cost by 10\% and 8\%, respectively. 
The recommended storage hierarchy outperforms its NVM-SSD counterpart by 19\%.
This is because the DRAM buffer reduces the time spent on NVM read operations 
by 63\%.

The optimal storage system for the \benchVoter workload consists of 4~GB
DRAM and 128~GB NVM, as shown in~\cref{fig:price-voter}. 
While executing this workload, the buffer manager frequently flushes dirty
blocks to durable storage. In the absence of NVM, the buffer manager 
spends more time flushing data to SSD. So, the performance/price number 
on a similarly priced 128~GB DRAM-SSD system is 16$\times$ lower than its
NVM-based counterpart.

On the \benchAuctionmark workload, as shown in~\cref{fig:price-auctionmark}, 
a DRAM-SSD system consisting of 8~GB DRAM delivers the highest performance/price
number. It delivers 2.9$\times$ lower throughput compared to a 3.4$\times$
higher priced NVM-SSD system with 1~TB NVM. 
We attribute this to the workload's smaller working set that fits in the DRAM
buffer. So, the utility of the NVM buffer is not as prominent on this workload.
Adding a 4~GB DRAM buffer on top of NVM-SSD hierarchy does not improve
performance on the \benchAuctionmark workload. Instead, it reduces throughput by
6\%. The I/O overhead associated with data migration between DRAM and NVM
overrides the utility of caching data on DRAM.

For the \benchChbenchmark workload, the results in~\cref{fig:price-chbenchmark}
show that the maximal performance/price number is delivered by a DRAM-NVM-SSD
system with 4~GB DRAM and 1~TB NVM. Adding a 4~GB DRAM buffer on top of NVM 
increases throughput by 5\% on this workload. This is because it reduces time
spent on NVM operations by 11\%, thereby justifying the cost of data migration.
\\ \vspace{-0.05in}

\textbf{Impact of NVM latency:} 
We next examine the impact of NVM latency on the selection of storage
hierarchy. ~\cref{fig:price-tpcc-8} presents the results for the \benchTPCC
benchmark with the 8$\times$ latency configuration. The storage system that delivers 
the highest performance/price number consists of 16~GB DRAM and 512~GB NVM 
on top of SSD. The capacity of the NVM buffer has shrunk from 1~TB with the
2$\times$ latency configuration. This shows that the utility of the NVM buffer
has decreased due to slower NVM accesses.

The results in~\cref{fig:price} illustrate how the selection of a multi-tier
storage system for a given workload depends on the working set size, 
the frequency of persistent writes, the performance and cost characteristics of
NVM, and the system cost budget.

\subsection{Policy Comparison}
\label{sec:exps-comparison}

\begin{table}[t!]
    \centering
    {\small {

\newcolumntype{b}{X}
\newcolumntype{Y}{>{\centering\arraybackslash}X}
\begin{tabularx}{\columnwidth}{YYYYY}
\toprule
\textbf{Policy} & \textbf{$\mathcal{D}_{r}$} &
\textbf{$\mathcal{D}_{w}$} & \textbf{$\mathcal{N}_{r}$} &
\textbf{$\mathcal{N}_{w}$}  \\

\midrule

$\mathcal{A}$~\cite{renen18} & 1 & 1 & 0.01 & 0.5  \\ 
$\mathcal{B}$ & 0.01 & 0.01 & 0.2 & 1  \\
$\mathcal{C}$ & 0.01 & 0.01 & 0.2 & 0.5  \\
$\mathcal{D}$ & 0.01 & 0.01 & 0.2 & 0.3  \\

\bottomrule
\end{tabularx}
}}
    \caption{
    	\textbf{Policy Comparison:} List of buffer management policies.
    }
    \label{tab:policy}
\end{table}

We now compare a family of buffer management policies against the policy
presented in~\cite{renen18}.
As shown in ~\cref{tab:policy},
$\mathcal{A}$ consists of eager migration for DRAM ($\mathcal{D}_{r}$ = 1,
$\mathcal{D}_{w}$ = 1), and lazy migration for NVM during reads
($\mathcal{N}_{r}$ = 0.01) and writes ($\mathcal{N}_{w}$ = 0.5).
We construct policies $\mathcal{B}$, $\mathcal{C}$, and $\mathcal{D}$ based on
the data migration optimizations in~\cref{sec:exps-migration} 
to improve runtime performance and extend the lifetime of the NVM device.
These policies adopt lazy migration to DRAM ($\mathcal{D}_{r}$ = 0.01,
$\mathcal{D}_{w}$ = 0.01) and NVM during reads ($\mathcal{N}_{r}$ = 0.2), 
and differ in how they migrate data to NVM during writes ($\mathcal{N}_{w}$ =
[1, 0.5, 0.3]). They differ from $\mathcal{A}$ in two ways.
With $\mathcal{A}$, the buffer manager initially moves data to DRAM and stores
data evicted from DRAM on NVM. It bypasses NVM during writes to ensure that only
frequently referenced data is stored on NVM. 
In contrast, with the former policies, the buffer manager initially moves data
into NVM and lazily migrates it to DRAM. It frequently bypasses DRAM during
writes and directly persists data on NVM.

\begin{figure}[t!]
    \centering
    \fbox{\includegraphics[width=0.45\textwidth,trim=2.5cm 0.5cm 0.3cm 0.3cm]
            {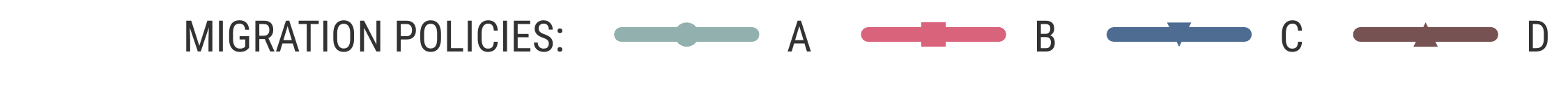}}
    \\[-0.1ex] 
    \subfloat[TPCC]{
        \includegraphics[width=0.22\textwidth]
                        {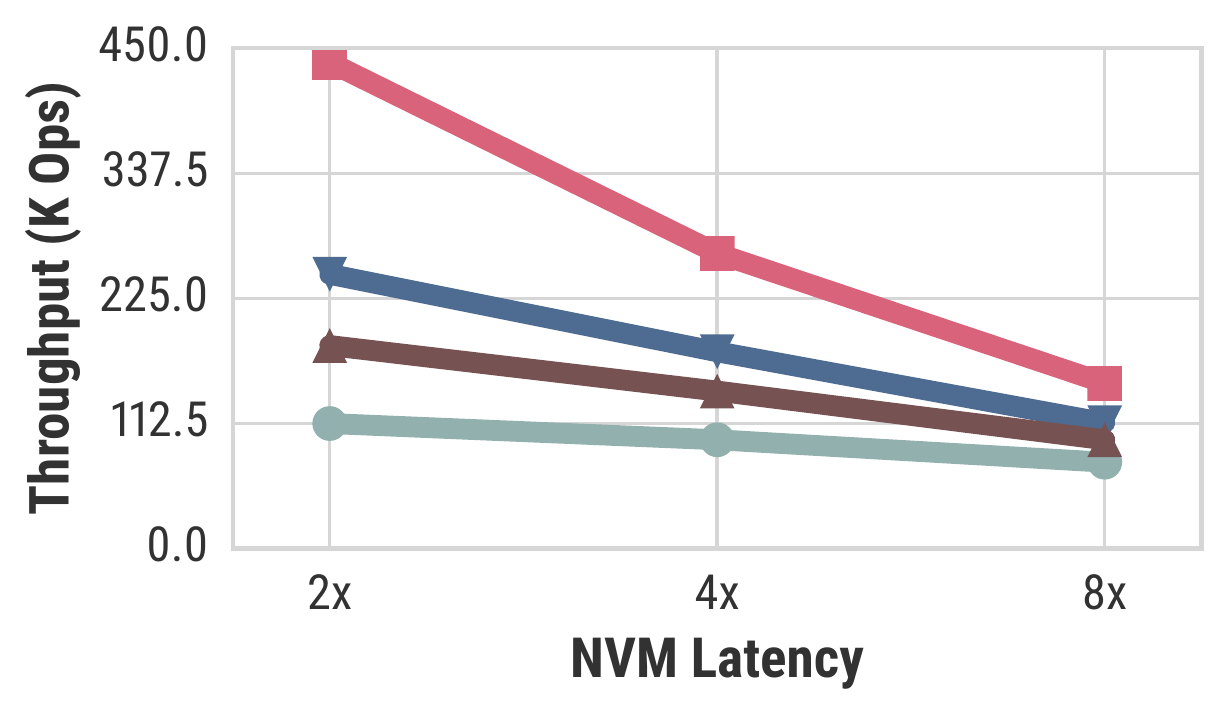}
        \label{fig:policy-tpcc}
    }
    \hfill
    \subfloat[VOTER]{
        \includegraphics[width=0.22\textwidth]
                        {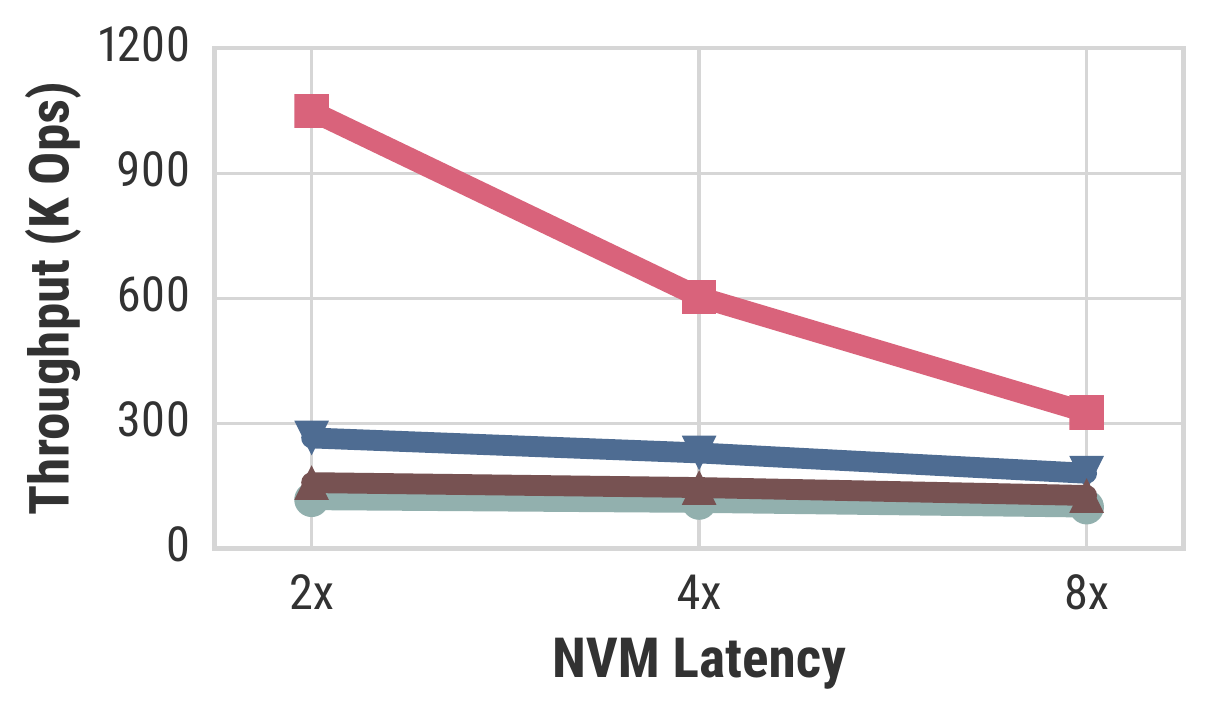}
        \label{fig:policy-voter}
    }                               
    \caption{
		\textbf{Impact of Policies on Runtime Performance:} The
		impact of different buffer management policies on runtime performance
		across different NVM latency configurations.
    }    
    \label{fig:policy}
\end{figure}

The results in~\cref{fig:policy} illustrate that $\mathcal{B}$ works 
well across different workloads. For the \benchTPCC workload shown
in~\cref{fig:policy-tpcc}, it outperforms $\mathcal{A}$ by 3.5$\times$.
The reasons for this are twofold. 
First, with $\mathcal{A}$, the buffer manager bypasses NVM during writes.
Although this scheme reduces the number of writes to NVM by 2$\times$, 
it increases the time spent by the buffer manager on SSD operations by
8.8$\times$. The former policy circumvents this problem by absorbing 
more writes on NVM. The buffer manager reclaims space in the NVM buffer by
discarding unmodified blocks.
Second, bypassing DRAM during reads ($\mathcal{D}_{r}$ = 0.01) reduces the data
migration overhead between NVM and DRAM and ensures that only frequently
referenced blocks are stored on DRAM.

\begin{figure}[t!]
    \centering
    \subfloat[TPCC]{
        \includegraphics[width=0.22\textwidth]
                        {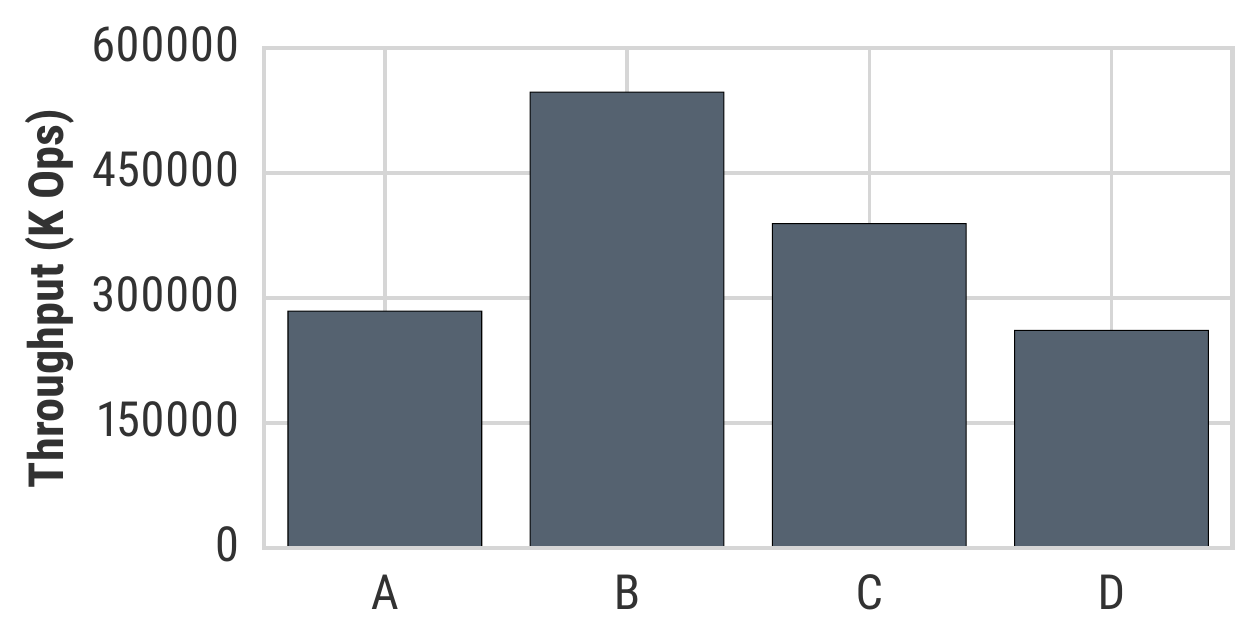}
        \label{fig:policy-write-tpcc}
    }
    \hfill
    \subfloat[VOTER]{
        \includegraphics[width=0.22\textwidth]
                        {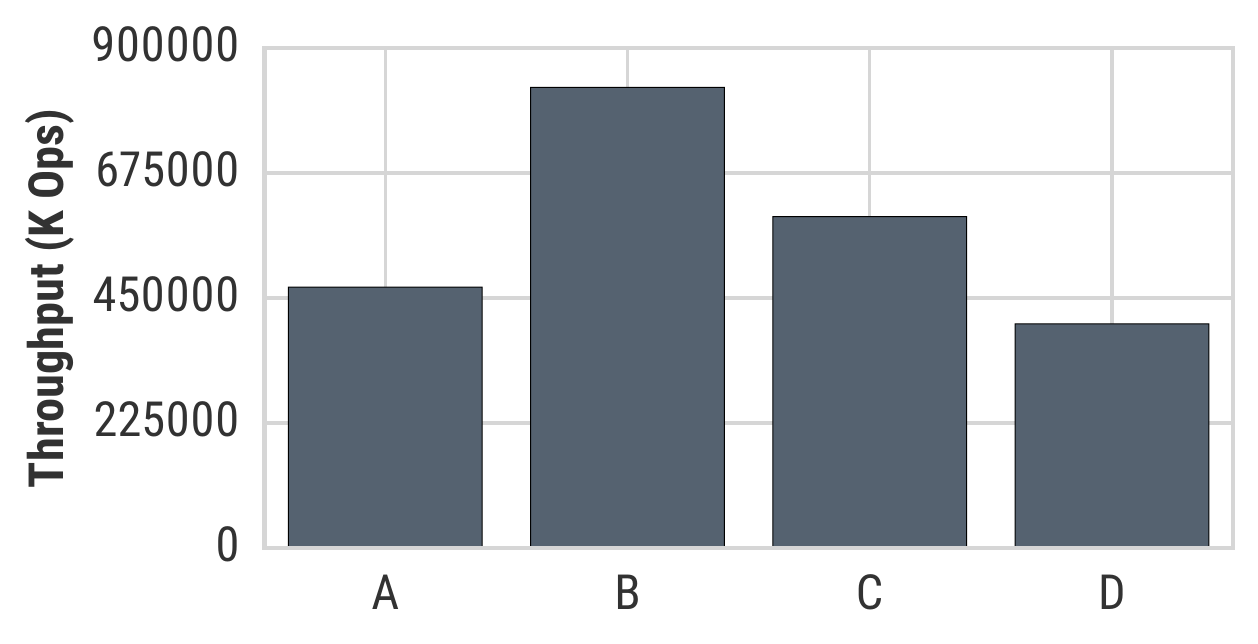}
        \label{fig:policy-write-voter}
    }                               
    \caption{
		\textbf{Impact of Policies on NVM Device Lifetime:} The
		impact of different policies on lifetime of NVM device.
    }    
    \label{fig:policy-write}
\end{figure}

The results in~\cref{fig:policy-voter} show the utility of eager migration
to NVM during writes. $\mathcal{B}$ outperforms $\mathcal{A}$ by 6.6$\times$ on
this workload. With the former policy, the buffer manager directly persists data
on NVM instead of first buffering it on DRAM. Since DRAM write latencies are
comparable to those of NVM, particularly on the 2$\times$ latency configuration,
bypassing DRAM during writes reduces the overall write latency, 
thereby improving runtime performance. 

The performance impact of NVM latency is not as prominent with $\mathcal{A}$.
The throughput only drops by 18\% when we transition from a 2$\times$ latency 
configuration to a 8$\times$ configuration. This is because lazy migration to
NVM increases the time spent on SSD operations, thereby reducing the impact of
slower NVM operations. 
\\ \vspace{-0.05in}

\textbf{Impact on NVM Device Lifetime:}
With $\mathcal{B}$, the buffer manager performs 2$\times$ more writes to NVM
than with $\mathcal{A}$. This shrinks the lifetime of NVM devices with limited
write-endurance~\cite{raoux08}. To circumvent this problem, we construct 
policies $\mathcal{C}$ and $\mathcal{D}$, which lazily migrate data to NVM 
during writes (i.e., $\mathcal{N}_{w}$ < 1).

The results in~\cref{fig:policy-write} illustrate the impact of these
policies on device lifetime. In comparison to $\mathcal{B}$, 
the number of writes to NVM on the \benchTPCC workload drops by 1.4$\times$ and
2.1$\times$ with $\mathcal{C}$ and $\mathcal{D}$, respectively. 
These policies outperform $\mathcal{A}$ by 2.2$\times$ and 1.6$\times$,
respectively. The buffer manager picks among these policies depending on the
write-endurance characteristics of the NVM device.

The results in~\cref{fig:policy,fig:policy-write} illustrate how a combination
of data migration optimizations presented in~\cref{sec:buffer-management}
maximizes both runtime performance and device lifetime.

\section{Related Work}
\label{sec:related-work}

We now discuss the previous research on NVM, especially in the context of buffer
management in DBMSs and file-systems.
\\ \vspace{-0.05in}

\textbf{NVM-Aware Buffer Management in DBMSs:}
Renen et al. present a NVM-aware multi-tier  buffer manager that eagerly
migrates data from SSD to DRAM~\cite{renen18}. When a page is evicted from DRAM,
the buffer manager considers admitting it into the NVM buffer. The key idea is
to only admit recently referenced pages. The buffer manager maintains an admission 
queue to keep track of pages considered for admission and only admits pages that
were recently denied admission. While this buffer management policy works well in
their target storage hierarchy, it does not generalize to other NVM
technologies, storage hierarchies, and workloads. We introduce a taxonomy of 
data migration optimizations that subsumes the specific scheme adopted in their
system. We study how the optimal policy depends on workload and storage system
characteristics and address the problem of designing a multi-tier storage system.

SOFORT~\cite{oukid14} is a hybrid storage engine that targets a two-tier storage
system with DRAM and NVM. The engine stores the primary copy of the data on NVM,
and supports the maintenance of auxiliary data structures on DRAM and NVM.
FOEDUS is a scalable OLTP engine designed for a two-tier storage system with
DRAM and NVM~\cite{kimura15}. It is based on the dual page primitive
that points to a pair of logically equivalent pages, a mutable volatile page in
DRAM containing the latest changes, and an immutable snapshot page on NVM.
Arulraj et al. compare three storage management architectures for an NVM-only
system and demonstrate that in-place updates architecture maximizes performance
and minimizes the wear on the NVM device~\cite{arulraj15}. Unlike these systems,
this paper focuses on managing and designing multi-tier storage hierarchy with
DRAM, NVM, and SSD.
\\ \vspace{-0.05in}

\textbf{NVM-aware Buffer Management in File Systems:}
Beyond DBMSs, researchers have explored using NVM in file-systems.
BPFS uses a variant of shadow paging on NVM to support atomic fine-grained
updates by relying on a special hardware instruction that ensures ordering
between writes in different epochs~\cite{condit09}.
PMFS is another filesystem from Intel Labs that is designed for byte-addressable
NVM~\cite{dulloor14}.  It relies on a write-ahead logging for meta-data and uses
shadow paging for data.
EXT4 DAX extends the EXT4 file system to support direct mapping of NVM
by bypassing the buffer cache~\cite{lwn14}.
Aerie provides direct access for file data I/O using user-level leases for NVM
updates~\cite{volos14}. 
NOVA is a novel per-inode log-structured file system that provide synchronous 
file system semantics on NVM, but requires system calls for every
operation~\cite{xu16}.
F2FS is an SSD-optimized log-structured  file system that sorts data to reduce 
file system write amplification~\cite{lee15}. 

Strata is a cross-media file system that supports performance-isolated access to
NVM using a per-application log by efficiently operating on SSDs and
HDDs~\cite{kwon17}. This system is optimized for a specific NVM technology that 
is 2$\times$ slower than DRAM. So it does not cache NVM-resident data on DRAM.
For the same reason, it bypasses DRAM while performing synchronous write operations.
While this buffer management policy works well in their target environment, 
it does not generalize to other NVM technologies, storage hierarchies, and
workloads. We tackle the buffer management problem within the context of a DBMS.
Operating inside a DBMS allows us to support, and requires us to handle, a
broader class of application access patterns.
\\ \vspace{-0.05in}

\textbf{Buffer Management in Systems without NVM:} 
Before the advent of NVM technologies, researchers have studied multi-tier
buffer management in storage systems without NVM.
FlashStore is a key-value store that uses an SSD as a fast cache between DRAM and
HDD and minimizes the number of SSD accesses~\cite{debnath10}.
RAMCloud is a sharded data storage system that uses disk as a back up for data
stored on replicated DRAM~\cite{ousterhout15}. It improves the DRAM utilization
by employing a log-structured design on both DRAM and disk~\cite{ongaro11}. 
Nitro is an SSD caching system that relies on data compression and deduplication
to maximize storage utilization~\cite{li14}.
RIPQ is a novel caching layer that shrinks write amplification by using the
local SSD as a read-only cache for remote storage~\cite{tang15}. 
Unlike these systems, this paper focuses on NVM-aware buffer management.
\\ \vspace{-0.05in}

\section{Conclusion}
\label{sec:conclusion}

This paper presented techniques for managing and designing a multi-tier storage
hierarchy comprising of DRAM, NVM, and SSD. We introduced a taxonomy for
NVM-aware data migration optimizations and discussed how the buffer management
policy can be synthesized based on the performance requirements and write
endurance characteristics of NVM.
We presented an adaptation mechanism in the buffer manager that achieves a 
near-optimal policy for an arbitrary workload and storage hierarchy without
requiring any manual tuning. We developed a recommendation system for designing
a multi-tier storage hierarchy for a target workload and system cost budget. 
Our results demonstrate that the NVM-aware buffer manager and storage system
designer improve system throughput and reduce system cost across different
transaction and analytical processing workloads.

\clearpage

\bibliographystyle{abbrvnat}
{\small
\raggedright
\bibliography{migration}}

\appendix

\section{NVM Hardware Emulator}
\label{sec:emulator}

The hardware emulator is a dual-socket system equipped with Intel Xeon E5-4620
CPUs (2.6 GHz), each with eight cores and a 20 MB L3 cache. Each processor
supports four DDR3 channels with two DIMMs per channel. PMEP reserves half of
the memory channels on each processor for emulated NVM while using the rest for
regular memory. The emulator's custom BIOS firmware partitions the physical
memory address space into separate address spaces for DRAM and emulated NVM.

NVM technologies have higher read and write latency than DRAM. PMEP emulates 
the latency for the NVM partition using custom CPU microcode.
The microcode estimates the additional cycles that the CPU would have to
wait if DRAM is replaced by slower NVM and then stalls the CPU for those cycles.
The accuracy of the latency emulation model has been validated by comparing the
performance of applications on emulated NVM and slower NUMA
memory~\cite{dulloor14}. The emulator throttles the write bandwidth
by limiting the number of DDR operations performed per microsecond.

The buffer manager uses the filesystem interface exported by the emulator. 
This allows the buffer manager to use the POSIX filesystem interface to
read/write data to files stored on NVM. 
This interface is implemented by the \textit{persistent memory filesystem}, a
special filesystem optimized for NVM~\cite{pmfs}. 
Normally, in a block-oriented filesystem, file I/O requires two copies; one
involving the block device and another involving the user buffer. The emulator's
optimized filesystem, however, requires only one copy between the file and the
user buffers.

\section{Additional Related Work}
\label{sec:additional-related-work}

\textbf{In-memory DBMSs:} 
Prior research has shown that there is significant overhead associated with
buffer management in a DBMS. 
When all the data fits in main memory, the cost of maintaining a buffer pool is
nearly one-third of all the CPU cycles used by the DBMS~\cite{harizopoulos08}.
This is because the buffer manager must keep track of meta-data about pages in
the pool to enforce the buffer replacement policy and synchronize
concurrent accesses from different threads to the pool.
The overhead associated with managing disk-resident
data has given rise to a class of new in-memory DBMSs that manage the entire database in main
memory and do not contain a buffer pool~\cite{dewitt84,memsql16,timesten07}.

In-memory DBMSs provide better throughput and lower latency than disk-based
DBMSs on OLTP applications due to this main memory orientation~\cite{kallman08}.
The fundamental limitation of in-memory DBMSs, however, is that they can deliver
this improved performance only when the database is smaller than the amount of
DRAM available in the system. If the dataset grows larger than the memory capacity, 
then the operating system will start to page virtual memory, and main memory
accesses will cause page faults~\cite{Stonebraker1981}.
The execution of transactions is stalled until the pages are retrieved from
non-volatile storage. The performance of an in-memory DBMS drops by up to
66\% when the dataset exceeds the memory capacity, even if the working set fits
in memory~\cite{stoica13}.

\textbf{Larger-than-Memory Data Management:}
Several techniques have been proposed to improve the performance of in-memory
DBMSs while operating on larger-than-memory databases~\cite{Ma2016}. These
techniques exploit the skewed access patterns observed in modern database
applications. In these workloads, certain hot data tuples are 
accessed more frequently than other cold tuples. 
While handling such workloads, it is advantageous to cache the hot data in
memory since it is likely to be modified during this period.
But then once the age of particular tuple crosses some threshold, 
the buffer manager can migrate the cold tuple out to cheaper secondary storage.
With this data migration technique, the DBMS can still deliver high performance
for transactions that operate on hot in-memory tuples while still being able to
access the cold data if needed at a later point in time. 
This paper generalizes these buffer management techniques to a multi-tier
storage hierarchy.

\end{document}